%% file: HIG-17-033_temp.tex
\pdfoutput=1

\documentclass[11pt,twoside,a4paper,cmspaper,final,collab]{cms-tdr}
\def\svnVersion{c2e7d6a-D}\def\svnDate{2020/03/09}\def\cmsCernNoTag{CERN-EP-2019-230}\def\cmsCernDate{\today}\def\cmsMessage{"Published in the Journal of High Energy Physics as \href{http://dx.doi.org/10.1007/JHEP03(2020)034}{\doi{10.1007/JHEP03(2020)034}.}"}

\begin{document}\cmsNoteHeader{HIG-17-033}

\hyphenation{had-ron-i-za-tion}
\hyphenation{cal-or-i-me-ter}
\hyphenation{de-vices}
\RCS$Revision: 494869 $
\RCS$HeadURL: svn+ssh://svn.cern.ch/reps/tdr2/papers/HIG-17-033/trunk/HIG-17-033.tex $
\RCS$Id: HIG-17-033.tex 494869 2019-05-24 12:46:12Z dmoran $

\newlength\cmsTabSkip\setlength\cmsTabSkip{3pt}
\providecommand{\cmsTable}[1]{\resizebox{\textwidth}{!}{#1}}

\newcommand{\ra}{\ensuremath{\to}\xspace}
\newcommand{\W}{\ensuremath{\PW}\xspace}
\newcommand{\h}{\ensuremath{\Ph}\xspace}
\newcommand{\hsm}{\ensuremath{\Ph(125)}\xspace}
\newcommand{\bq}{\ensuremath{\cPqb}\xspace}
\newcommand{\WW}{\ensuremath{\W\W}\xspace}
\newcommand{\WZ}{\ensuremath{\W\PZ}\xspace}
\newcommand{\ZZ}{\ensuremath{\PZ\PZ}\xspace}
\newcommand{\pp}{\ensuremath{\Pp\Pp}\xspace}
\newcommand{\VVV}{\ensuremath{\PV\PV\PV}\xspace}
\newcommand{\Wjets}{\ensuremath{\W\text{+jets}}\xspace}
\newcommand{\LN}{\ensuremath{\ell\nu}\xspace}
\newcommand{\FL}{\ensuremath{2\ell 2\nu}\xspace}
\newcommand{\SL}{\ensuremath{\ell\nu 2\PQq}\xspace}
\newcommand{\ggf}{\ensuremath{\Pg\Pg \text{F}}\xspace}
\newcommand{\emu}{\ensuremath{\Pe\Pgm}\xspace}
\newcommand{\wqqb}{\ensuremath{\cPq\cPaq'}\xspace}
\newcommand{\qqb}{\ensuremath{\cPq\cPaq}\xspace}
\newcommand{\qqWWqq}{\ensuremath{\cPq\cPq \ra \cPq\cPq \WW}\xspace}
\newcommand{\qqWW}{\ensuremath{\qqb \ra \WW}\xspace}
\newcommand{\ggWW}{\ensuremath{\Pg\Pg \ra \WW}\xspace}
\newcommand{\gghsm}{\ensuremath{\Pg\Pg \ra \hsm}\xspace}
\newcommand{\tautau}{\ensuremath{\Pgt^{+} \Pgt^{-}}\xspace}
\newcommand{\elel}{\ensuremath{\Pep \Pem}\xspace}
\newcommand{\mumu}{\ensuremath{\Pgmp \Pgmm}\xspace}
\newcommand{\fvbf}{\ensuremath{f_\text{VBF}}\xspace}
\newcommand{\mmodp}{\ensuremath{m_{\h}^\text{mod+}}\xspace}
\newcommand{\mmodh}{\ensuremath{M_{\h}^{125}}\xspace}
\newcommand{\mmoda}{\ensuremath{M_{\h}^{125}\text{(alignment)}}\xspace}
\newcommand{\mmodc}{\ensuremath{M_{\h}^{125}(\PSGc)}\xspace}
\newcommand{\mmodt}{\ensuremath{M_{\h}^{125}(\PSGt)}\xspace}
\newcommand{\ma}{\ensuremath{m_{\PSA}}\xspace}
\newcommand{\mH}{\ensuremath{m_{\PH}}\xspace}
\newcommand{\mHpm}{\ensuremath{m_{\PHpm}}\xspace}
\newcommand{\mX}{\ensuremath{m_{\PX}}\xspace}
\newcommand{\tanbeta}{\ensuremath{\tan\beta}\xspace}
\newcommand{\ptww}{\ensuremath{\pt^{\WW}}\xspace}
\newcommand{\mww}{\ensuremath{m_{\WW}}\xspace}
\newcommand{\pz}{\ensuremath{p_z}\xspace}
\newcommand{\ptll}{\ensuremath{\pt^{\ell\ell}}\xspace}
\newcommand{\ptl}{\ensuremath{\pt^{\ell}}\xspace}
\newcommand{\ptljj}{\ensuremath{\pt^{\ell \mathrm{jj}}}\xspace}
\newcommand{\ptvecll}{\ensuremath{\ptvec^{\ell\ell}}\xspace}
\newcommand{\ptvecl}{\ensuremath{\ptvec^{\ell}}\xspace}
\newcommand{\ptvecljj}{\ensuremath{\ptvec^{\ell \mathrm{jj}}}\xspace}
\newcommand{\ptw}{\ensuremath{\pt^{\W}}\xspace}
\newcommand{\mll}{\ensuremath{m_{\ell\ell}}\xspace}
\newcommand{\mthll}{\ensuremath{\mT^{\ell\ell}}\xspace}
\newcommand{\mthljj}{\ensuremath{\mT^{\ell \mathrm{jj}}}\xspace}
\newcommand{\dpll}{\ensuremath{\Delta\phi^{\ell\ell}}\xspace}
\newcommand{\dpl}{\ensuremath{\Delta\phi^{\ell}}\xspace}
\newcommand{\dpljj}{\ensuremath{\Delta\phi^{\ell \mathrm{jj}}}\xspace}
\newcommand{\mllnn}{\ensuremath{m_\text{reco}}\xspace}
\newcommand{\mtw}{\ensuremath{\mT^{\ell}}\xspace}
\newcommand{\mj}{\ensuremath{m_\mathrm{J}}\xspace}
\newcommand{\mjj}{\ensuremath{m_\mathrm{jj}}\xspace}
\newcommand{\tauo}{\ensuremath{\tau_{1}}\xspace}
\newcommand{\taut}{\ensuremath{\tau_{2}}\xspace}
\newcommand{\tauto}{\ensuremath{\tau_{21}}\xspace}
\newcommand{\RNum}[1]{\uppercase\expandafter{\romannumeral #1\relax}}
\newcommand{\chisq}{\ensuremath{\chi^{2}}\xspace}
\newcommand{\cosba}{\ensuremath{\cos(\beta-\alpha)}\xspace}
\newcommand{\sqrts}{\ensuremath{\sqrt{s}}\xspace}

\cmsNoteHeader{HIG-17-033}
\title{Search for a heavy Higgs boson decaying to a pair of \W bosons in proton-proton collisions at $\sqrt{s} = 13\TeV$}
\author{ \LARGE The CMS Collaboration}
\date{\today}

\abstract{A search for a heavy Higgs boson in the mass range from 0.2 to 3.0\TeV, decaying to a pair of \W bosons, is presented. The analysis is based on proton-proton collisions at $\sqrt{s} = 13\TeV$ recorded by the CMS experiment at the LHC in 2016, corresponding to an integrated luminosity of 35.9\fbinv. The \W boson pair decays are reconstructed in the \FL and \SL final states (with $\ell = \Pe$ or \Pgm). Both gluon fusion and vector boson fusion production of the signal are considered. Interference effects between the signal and background are also taken into account. The observed data are consistent with the standard model (SM) expectation. Combined upper limits at 95\% confidence level on the product of the cross section and branching fraction exclude a heavy Higgs boson with SM-like couplings and decays up to 1870\GeV.
Exclusion limits are also set in the context of a number of two-Higgs-doublet model formulations, further reducing the allowed parameter space for SM extensions.
}

\hypersetup{pdfauthor={CMS Collaboration},pdftitle={Search for a heavy Higgs boson decaying to a pair of W bosons in proton-proton collisions at sqrt(s) = 13 TeV},pdfsubject={CMS},pdfkeywords={CMS, physics, Heavy Higgs, WW final states}}

\maketitle

\section{Introduction}\label{sec:intro}

The discovery of the standard model (SM) Higgs boson, with a mass close to 125\GeV, by the CERN LHC experiments ATLAS and CMS in 2012~\cite{Aad:2012tfa, Chatrchyan:2012xdj, Chatrchyan:2013lba} represents a major advancement in particle physics.
Studies of the new particle have so far shown consistency with the SM Higgs mechanism predictions~\cite{Aad:2015mxa, Aad:2019mbh, Aaboud:2018ezd, Aaboud:2018puo, Aaboud:2017vzb, Khachatryan:2016vau, Khachatryan:2016tnr, Sirunyan:2017tqd, Sirunyan:2019twz, Sirunyan:2018sgc, Sirunyan:2018koj, Sirunyan:2019nbs}.
Throughout this paper, the observed SM Higgs boson is denoted as \hsm.
In order to determine whether the SM gives a complete description of the Higgs sector, precise measurements of the \hsm coupling strengths, CP structure and kinematic distributions are required~\cite{Englert:2014uua, Plehn:2001nj, Anderson:2013afp, Bishara:2016jga, Grazzini:2016paz}.
A complementary strategy involves the search for an additional Higgs boson, denoted \PX, whose existence would prove the presence of beyond the SM (BSM) physics in the form of a non minimal Higgs sector~\cite{Barger:2007im, Branco:2011iw}. The search for an additional scalar resonance in the full mass range accessible at the LHC remains one of the main objectives of the experimental community.

The search for a high-mass Higgs boson has been performed at ATLAS~\cite{Aad:2015agg, Aad:2015kna, Aaboud:2017gsl, Aaboud:2017rel} and CMS~\cite{Khachatryan:2015cwa, Sirunyan:2018qlb} in a number of final states, using proton-proton (\pp) collisions at centre-of-mass energies (\sqrts) of 7, 8 and 13\TeV, with no significant excess observed.
For Higgs boson masses above 200\GeV one of the most sensitive channels is the decay to a pair of \W bosons~\cite{Branco:2011iw}.
In this analysis, a search is performed in the fully leptonic, \FL, and semileptonic, \SL, \WW decay channels (with $\ell = \Pe$ or \Pgm) using \pp collisions recorded at $\sqrts = 13\TeV$ by the CMS experiment in 2016, corresponding to an integrated luminosity of 35.9\fbinv.

The fully leptonic channel has a clear signature of two isolated leptons and missing transverse momentum (\ptmiss), due to the neutrinos escaping detection.
For the semileptonic channel, the leptonically decaying boson is reconstructed as a single isolated lepton and \ptmiss.
The hadronically decaying boson may be sufficiently boosted that its decay products are contained in a single merged jet.
Jet substructure techniques are used to identify merged jets with two well defined subjets and to determine the merged jet mass, helping to discriminate vector boson hadronic decays from other jets.
When the \W boson hadronic decay products are resolved, it may be reconstructed using two quark jets (a dijet).
The search is performed in a wide mass range from 0.2 up to 3.0\TeV.
Events are categorized to enhance the sensitivity to the gluon fusion (\ggf) and vector boson fusion (VBF) Higgs boson production mechanisms.

A signal interpretation in terms of a heavy Higgs boson with SM-like couplings and decays is performed.
This is motivated by BSM models in which the \hsm mixes with a heavy electroweak singlet, resulting in an additional resonance at high-mass with couplings similar to those of the SM Higgs boson~\cite{Barger:2007im}.
The signal model includes a detailed simulation of the interference between the \PX signal, the \hsm off-shell tail, and the \WW background~\cite{Kauer:2015hia}.
A number of hypotheses for the relative contribution of \ggf and VBF production are investigated.

Additional interpretations based on a number of two-Higgs-doublet models (2HDMs)~\cite{Branco:2011iw} are performed.
The 2HDM, which introduces a second scalar doublet, is incorporated in supersymmetric~\cite{Martin:1997ns} and axion~\cite{Kim:1986ax} models, and may introduce additional sources of explicit or spontaneous CP violation that could explain the baryon asymmetry of the Universe~\cite{Cline:2011mm}.
As will be discussed in Section~\ref{sec:anastr}, the measured properties of the \hsm set strong constraints on the decay of a heavy Higgs boson to vector bosons in the context of 2HDMs.

This paper is organized as follows: in Section~\ref{sec:cms}, a brief description of the CMS detector is provided; Section~\ref{sec:mcdata} gives a description of the data and Monte Carlo (MC) simulated samples
used in the analysis; Section~\ref{sec:evreco} provides a description of the event reconstruction; Section~\ref{sec:anastr} contains an overview of the signal models considered;
in Section~\ref{sec:sel}, the event selection and categorization are discussed; Section~\ref{sec:bkg} explains the estimation of the SM backgrounds;
the signal extraction procedure and the systematic uncertainties affecting the analysis are presented in Section~\ref{sec:sys}; the results are presented in Section~\ref{sec:results}.
Finally, results are summarized in Section~\ref{sec:conclusions}.

\section{The CMS detector}\label{sec:cms}

The CMS detector, described in detail in Ref.~\cite{Chatrchyan:2008zzk}, is a
multipurpose apparatus designed to study high transverse momentum (\pt)
physics processes in \pp and heavy-ion collisions.
A superconducting solenoid occupies its central region, providing a magnetic
field of 3.8\unit{T} parallel to the beam direction. Charged-particle
trajectories are measured by the silicon pixel and strip trackers, which
cover a pseudorapidity region of $\abs{\eta} < 2.5$.
A crystal electromagnetic calorimeter (ECAL), and
a brass and scintillator hadron calorimeter surround the tracking volume
and cover $\abs{\eta} < 3$. The steel and quartz-fiber Cherenkov hadron forward (HF)
calorimeter extends the coverage to $\abs{\eta} < 5$. The muon system consists of gas-ionization
detectors embedded in the steel flux return yoke outside the solenoid, and
covers $\abs{\eta} < 2.4$. The first level of the CMS trigger system~\cite{Khachatryan:2016bia},
composed of custom hardware processors, is designed to select the most
interesting events in less than 4\mus, using information from the
calorimeters and muon detectors. The high-level trigger processor farm
further reduces the event rate to 1\unit{kHz} before data storage.

\section{Data and simulated samples}\label{sec:mcdata}

The events used to study the \SL final state are selected by high-level trigger algorithms that require the presence of one electron with $\pt > 25\GeV$ and $\abs{\eta} < 2.1$ passing tight identification and isolation requirements,
or one muon with $\pt > 24\GeV$ and $\abs{\eta} < 2.4$ passing loose identification and isolation requirements.
The trigger efficiency for \SL signal events passing the offline event selection is about 93\%.
Both single-lepton and dilepton triggers are used to select events to study the \FL final state.
In addition to the single-lepton triggers described, the \FL final state events are also selected by a trigger which requires one electron outside the central region ($2.1 < \abs{\eta} < 2.5$) with $\pt > 27\GeV$.
The dilepton triggers require the presence of two leptons passing relatively loose identification and isolation requirements.
For the dielectron (dimuon) trigger, the \pt thresholds are 23 (17)\GeV for the leading and 12 (8)\GeV for the subleading electrons (muons).
For the different-flavour dilepton trigger, the \pt thresholds are either 8\GeV
for the muon and 23\GeV for the electron, or 23\GeV for the muon and 12\GeV for the electron.
The overall trigger efficiency for the combination of the single-lepton and dilepton triggers for \FL signal events passing the offline event selection is larger than 99\%.

Several event generators are used to optimize the analysis and estimate the yields of signal and background events, as well as the associated systematic uncertainties.
The heavy Higgs boson signal samples are generated in the \ggf and VBF production modes at next-to-leading order (NLO) in quantum chromodynamics (QCD) using \POWHEG v2~\cite{Nason:2004rx,Frixione:2007vw,Alioli:2010xd,Bagnaschi:2011tu,Nason:2009ai}, for a number of masses ranging from 0.2 to 3.0\TeV.
The resonance width is set according to the SM Higgs boson expectation for signal masses up to 1\TeV.
For signal masses higher than 1\TeV the width is set to half the resonance mass, which approximately corresponds to the SM Higgs boson prediction at 1\TeV.
The decay of the signal to a pair of \W bosons is simulated with \textsc{JHUGen}~v6.2.8~\cite{Gao:2010qx, Bolognesi:2012mm}.
The simulated signal samples are normalized using cross sections and decay rates computed by the LHC Higgs Cross Section Working Group~\cite{deFlorian:2016spz}.

The \Wjets process is produced at NLO with the \MGvATNLO v2.2.2 event generator~\cite{Alwall:2014hca}, using the FxFx merging scheme~\cite{Frederix:2012ps}
between the jets from matrix element calculations and parton showers (PS), and scaled to the next-to-NLO (NNLO) cross section computed using \FEWZ v3.1~\cite{Li:2012wna}.

Single top quark and \ttbar processes are generated at NLO using \POWHEG~\cite{Re:2010bp, Frixione:2007nw} and \MGvATNLO.
The cross sections of the different single top quark processes are calculated at NLO~\cite{Kant:2014oha}, while the \ttbar cross section is computed at NNLO, with next-to-next-to-leading-logarithmic soft-gluon
resummation~\cite{Czakon:2011xx}.

The \WW diboson continuum background is simulated in a number of ways.
The production of \WW via \qqb (\qqWW) is generated using \POWHEG~\cite{Melia:2011tj} and \MGvATNLO at NLO, \WW production via gluon fusion (\ggWW) is generated using \MCFM v7.0~\cite{Campbell:2013wga} at leading order (LO), while a \WW plus two jets (\qqWWqq) sample is produced with \MGvATNLO at LO.
The cross section used for normalizing the \WW processes produced via \qqb is computed at NNLO ~\cite{Gehrmann:2014fva}.
For the \ggWW process, the difference between LO and NLO cross sections is large; a scale factor of 1.4 is theoretically calculated~\cite{Caola:2015rqy} and applied to the cross section prediction from \MCFM.
In order to suppress the top quark background processes, the \FL analysis implements an event categorization based on jet multiplicity.
This approach spoils the convergence of fixed-order calculations of the \qqWW process and requires the use of dedicated resummation techniques for an accurate prediction of the differential distributions~\cite{Meade:2014fca,Jaiswal:2014yba}.
The simulated \qqWW events are therefore reweighted to reproduce the \ptww distribution from the \pt-resummed calculation.

Drell--Yan (DY) production of $\PZ/\Pgg^{*}$ is generated at NLO using \MGvATNLO and scaled to the NNLO cross section computed using \FEWZ.
Multiboson processes such as \WZ, \ZZ, and \VVV (\PV = \W, \PZ) are also simulated at NLO with \MGvATNLO.

The QCD multijet production background is generated with \PYTHIA 8.212~\cite{Sjostrand:2014zea}.
The QCD samples are enriched in events containing electrons or muons with dedicated filters.

All processes are generated using the NNPDF 3.0~\cite{Ball:2013hta,Ball:2011uy} parton distribution functions (PDFs), with the order matching that of the matrix element calculations.
All the event generators are interfaced with \PYTHIA for showering of partons and hadronization,
and to simulate the underlying event (UE) and multiple-parton interactions based on the CUET8PM1 tune (CUETP8M2T4 for \ttbar samples)~\cite{Khachatryan:2015pea}.
To estimate systematic uncertainties related to the choice of UE and multiple-parton interactions tune, \WW background samples are generated with two alternative tunes, which are representative of the uncertainties in the tuning parameters. A systematic uncertainty associated with showering and hadronization is estimated by interfacing the same samples with the
\HERWIG{}++ 2.7 generator~\cite{Richardson:2013nfo,Bahr:2008pv}, using the UE-EE-5C tune for the simulation of UE and multiple-parton interactions~\cite{Khachatryan:2015pea}.

For all processes, the detector response is simulated using a detailed description of the CMS detector, based on the \GEANTfour package~\cite{Agostinelli:2002hh}.
Additional \pp interactions simulated with \PYTHIA are overlaid on the event of interest to reproduce the number of interactions occurring simultaneously within the same bunch crossing (pileup) measured in data.

\section{Event reconstruction}\label{sec:evreco}

The particle-flow (PF) algorithm~\cite{Sirunyan:2017ulk} is used to reconstruct
the observable particles in the event. Clusters of energy deposits measured by
the calorimeters, charged particle tracks identified in the central tracking
system, and the muon detectors, are combined to reconstruct individual particles (PF candidates).

If more than one vertex is reconstructed, the vertex with the largest value of summed physics-object $\pt^2$ is taken to be the primary \pp interaction vertex.
The physics objects are those returned by a jet finding algorithm~\cite{Cacciari:2008gp, Cacciari:2011ma} applied to all charged tracks assigned to the
vertex, and the associated missing transverse momentum, computed as the negative vectorial sum of the \pt of those jets.

Electrons are reconstructed from a combination of the deposited energy of the ECAL clusters associated with the track reconstructed from the measurements determined by the inner tracker, and the energy sum of all photons spatially compatible with being bremsstrahlung from the electron track~\cite{Khachatryan:2015hwa}.
The electron candidates are required to have $\abs{\eta} < 2.5$. Additional requirements are applied to reject electrons originating from photon
conversions in the tracker material or jets mis-reconstructed as electrons.
Electron identification criteria rely on observables sensitive to the bremsstrahlung along the
electron trajectory, the geometrical and momentum-energy matching between the electron trajectory
and the associated supercluster, as well as ECAL shower shape observables and compatibility with the primary vertex.

Muon candidates are reconstructed by combining charged tracks in the muon detector with tracks reconstructed in the central tracking system~\cite{Sirunyan:2018fpa}.
They are required to have $\abs{\eta} < 2.4$. Identification criteria based on the number of hits in the
tracker and muon systems, the fit quality of the muon track, and the consistency of the trajectory with the primary vertex, are imposed on the muon candidates to reduce the misidentification rate.

Prompt leptons from electroweak interactions are usually isolated, whereas misidentified leptons and leptons from jets, are often accompanied by charged or neutral particles, and can arise from a secondary vertex.
Therefore leptons are required to be isolated from hadronic activity by requiring that the sum of the \pt of charged hadrons associated with the primary vertex,
and the \pt of neutral hadrons and photons, in a cone around the lepton of radius $\Delta R = \sqrt{\smash[b]{ (\Delta\phi)^2+(\Delta\eta)^2 }} = 0.4$ (where $\phi$ is the azimuthal angle in radians), is below a certain fraction of the lepton \pt.
To mitigate the effect of pileup on the isolation variable, a correction based on the mean event energy density~\cite{Cacciari:2007fd} is applied.

The jet reconstruction uses all PF candidates, except those charged candidates that are not associated with the primary vertex.
This requirement mitigates the effect of pileup for $\abs{\eta} < 2.5$.
Particle candidates are clustered using the anti-\kt algorithm~\cite{Cacciari:2008gp, Cacciari:2011ma} with a distance parameter of 0.4 (AK4) or 0.8 (AK8).
To reduce the residual pileup contamination from neutral PF candidates,
a correction based on jet median area subtraction~\cite{Cacciari:2007fd} is applied.
The jet energy is calibrated using both simulation and data following the technique described in~\cite{Chatrchyan:2011ds}.
Only AK4 jets with  $\pt > 30\GeV$ (20\GeV for \bq quark jets) and $\abs{\eta} < 4.7$ (2.4 for \bq quark jets) are considered.
The AK8 jets are required to have $\pt > 200\GeV$ and $\abs{\eta} < 2.4$.
Those AK4 (AK8) jets which overlap with a well identified and isolated lepton within a distance of $\Delta R = 0.4$ (0.8) are ignored.

The vector \ptvecmiss, whose magnitude is the \ptmiss in the event, is computed as the negative vectorial sum in the transverse plane of all the PF candidates momenta.
The \ptvecmiss is modified to account for the corrections to the energy scale of the jets described above.

A jet grooming procedure, which removes contributions from soft radiation and additional interactions, is used on the AK8 jets to help identify and discriminate between jets from Lorentz-boosted hadronic \W boson decays and jets from quarks and gluons.
First, the pileup mitigation corrections provided by the pileup per particle identification (PUPPI) algorithm~\cite{Bertolini2014} are applied.
The jets are then groomed by means of a modified mass drop algorithm~\cite{Dasgupta:2013ihk, Butterworth:2008iy}, known as the soft-drop algorithm~\cite{Larkoski:2014wba}, with parameters $\beta = 0$, $z_{\text{cut}} = 0.1$ and $R_0 = 0.8$.
The soft-drop mass (\mj) used in the \SL analysis is computed from the sum of the four-momenta of the jet constituents passing the grooming algorithm.

Discrimination between AK8 jets originating from \W boson decays and those originating from gluons and quarks is also achieved by using the $N$-subjettiness jet substructure variable~\cite{Thaler2011}.
This observable exploits the distribution of the jet constituents found in the proximity of the subjet axes to determine if the jet can be effectively subdivided into a number $N$ of subjets.
The generic $N$-subjettiness variable $\tau_N$ is defined using the \pt-weighted sum of the angular distance $\Delta R_{N,k}$ of the jet constituents $k$ with respect to the axis of the $N^\text{th}$ subjet:
\begin{equation}
\tau_N = \frac{1}{d_0} \sum_k p_{\mathrm{T},k} \min( \Delta R_{1,k}, \Delta R_{2,k}, \dots, \Delta R_{N,k} ).
\end{equation}
The normalization factor $d_0$ is defined as $d_0 = \sum_k p_{\mathrm{T},k} R_0$, with $R_0$ being the clustering parameter of the original jet.
The variable which best discriminates \W boson jets from those coming from quarks and gluons is the ratio of the 2- to 1-subjettiness: $\tauto = \taut/\tauo$.
The \tauto observable is calculated for the jet after applying the PUPPI algorithm corrections for pileup mitigation.

To identify jets coming from \bq quarks, a multivariate \bq tagging algorithm~\cite{Sirunyan:2017ezt} and the combined secondary vertex algorithm~\cite{Sirunyan:2017ezt} are used in the \FL and \SL analyses, respectively.
In both cases, the chosen working point corresponds to about 80\% efficiency for genuine \bq quark jets and to
a mistagging rate of about 10\% for light-flavour or gluon jets, and of about 40\% for \PQc quark jets.

For each event in the fully leptonic channel, at least two high-\pt lepton candidates
originating from the primary vertex are required. Opposite-charge dielectron pairs, dimuon pairs and electron-muon (\emu) pairs are accepted.
In the semileptonic channel, at least one high-\pt lepton candidate, and two AK4 jets or one AK8 jet, originating from the primary vertex are required.

\section{Signal models}\label{sec:anastr}

\begin{figure}[htbp]
\centering
{
\includegraphics[width=0.6\textwidth]{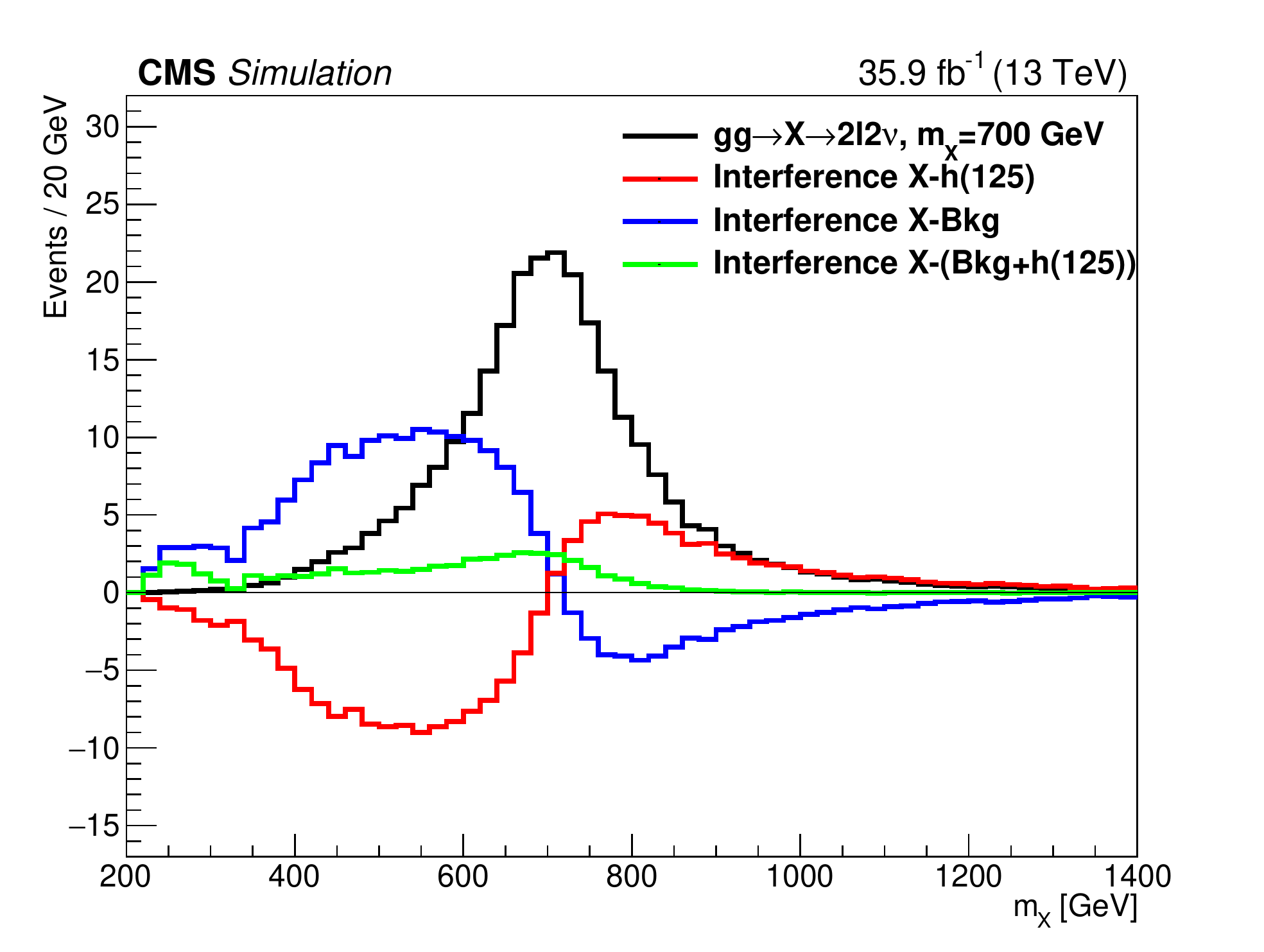}
}

\caption{Generator-level mass of a \ggf-produced 700\GeV signal (black line) normalized to the SM cross section and without considering interference effects.
        The effects of the interference of the signal with the \ggWW continuum and the \gghsm off-shell tail are shown, together with the total interference effect. }
    \label{fig:X300}
\end{figure}

A signal interpretation in terms of a heavy Higgs boson with SM-like couplings and decays is implemented in this analysis.
Both the \ggf and VBF production mechanisms are considered.
Due to the large expected width of the \PX resonance at high-mass, its interference with the \WW continuum and the \hsm off-shell tail becomes significant~\cite{Kauer:2015hia}.
The MELA matrix-element package~\cite{Gao:2010qx, Bolognesi:2012mm, Anderson:2013afp}, based on \textsc{JHUGen} for Higgs bosons, and on \MCFM for the continuum \WW background,
has been used to estimate the interference of high-mass \PX resonances with the \WW continuum and the \hsm.
The two sources of interference have opposite signs and partially cancel out with the size of the cancellation depending on the signal mass.
Figure~\ref{fig:X300} displays the generator-level mass distribution of a \ggf-produced 700\GeV signal and the effects of interference with the \ggWW continuum and \gghsm off-shell tail.
The interference effect is taken into account for both the \ggf and VBF production mechanisms.
A parameter \fvbf, which is the fraction of the VBF production cross section with respect to the total cross section, is included in the model and a number of hypotheses investigated.

An interpretation in the context of a general 2HDM is conducted.
Various formulations of the 2HDM predict different couplings of the two doublets to right-handed quarks and charged leptons:
in the Type-\RNum{1} formulation~\cite{Branco:2011iw}, all fermions couple to only one Higgs doublet;
in the Type-\RNum{2} formulation~\cite{Branco:2011iw}, the up-type quarks couple to a different doublet than the down-type quarks and leptons.
There are five physical Higgs bosons predicted: two CP-even neutral bosons \h and \PH; a neutral CP-odd boson \PSA; and two charged bosons \PHpm.
In most formulations of the 2HDM, \h corresponds to the \hsm, and \PH is an additional high-mass CP-even Higgs boson.
The 2HDM has two important free parameters, $\alpha$ and \tanbeta, which are the mixing angle and the ratio of the vacuum expectation values of the two Higgs doublets, respectively. The quantity \cosba is also of interest, as the coupling of the heavy Higgs boson \PH to two vector bosons is proportional to this factor.
In the alignment limit, which occurs at $\cosba = 0$, the properties of \h approach those of the SM Higgs boson, while the decay of \PH to vector bosons becomes heavily suppressed.
Based on the constraints given by the measurements of the \hsm couplings, the largest possible deviations of \cosba from 0 allowed are approximately 0.3 and 0.1 for the Type-\RNum{1} and -\RNum{2} scenarios respectively~\cite{Craig:2015jba, Chowdhury:2017aav}. Therefore the value of \cosba has been fixed to 0.1 for the 2HDM scenarios considered here.
In this way the measured properties of the \hsm are incorporated into the definition of the scenarios while still allowing for a non-negligible branching fraction for \PH to vector bosons.
In the limit that $\ma >> m_{\PZ}$, the masses of the \PH, \PSA, and \PHpm bosons become approximately degenerate. For simplicity it is assumed that $\mH = \ma = \mHpm$ for the 2HDM scenarios considered. 
The width of \PH has a dependence on \tanb, with relatively large widths predicted in comparison to both the SM widths and the experimental resolution for \tanbeta below ${\approx}0.2$ and \mH above ${\approx}400 \GeV$.
However, for the majority of the phase space explored the SM width assumption gives a reasonable approximation of the 2HDM predictions.

The minimal supersymmetric standard model (MSSM)~\cite{Fayet:1974pd, Fayet:1977yc}, which incorporates a Type-\RNum{2} 2HDM, is also considered.
At tree level, the whole phenomenology can be described using just two parameters.
By convention, these parameters are chosen to be \tanbeta and \ma, the mass of the pseudoscalar Higgs boson.
Beyond the tree level, the MSSM Higgs sector depends on additional parameters which enter via higher-order corrections in perturbation theory, and which are usually fixed to values motivated by experimental constraints and theoretical assumptions. The \mmodp~\cite{Carena:2013ytb} and hMSSM~\cite{Djouadi:2013vqa,Maiani:2013hud,Djouadi:2013uqa,Djouadi:2015jea} benchmark scenarios are defined by setting these parameters such that a wide range of the \ma-\tanbeta parameter space is compatible with the \hsm mass and production rate measurements at ATLAS and CMS. For the \mmodh, \mmoda, \mmodc, and \mmodt benchmark scenarios, a significant portion of the parameter space is consistent with the \hsm measurements and with limits from searches for supersymmetry particles and additional Higgs bosons at ATLAS and CMS using \pp collisions at $\sqrts = 7$, 8, and 13\TeV~\cite{Bahl:2018zmf}. The assumption of a SM width is a reasonable approximation for the MSSM scenarios considered, with relatively small widths predicted with respect to the experimental resolution for the majority of the phase space explored.

Model predictions for the MSSM scenarios are provided by the LHC Higgs Cross Section Working Group~\cite{deFlorian:2016spz}. 
The \ggf cross sections have been computed with \textsc{SusHi}~\cite{Harlander:2012pb, Harlander:2016hcx}, which includes NLO QCD corrections~\cite{Spira:1995rr}, NNLO QCD corrections for the top quark contribution in the effective theory of a heavy top quark~\cite{Harlander:2002wh,Anastasiou:2002yz,Ravindran:2003um} and electroweak effects by light quarks~\cite{Aglietti:2004nj,Bonciani:2010ms}. 
For most of the scenarios considered, NLO supersymmetric-QCD corrections~\cite{Harlander:2004tp,Harlander:2005rq,Degrassi:2010eu,Degrassi:2012vt} in expansions of heavy SUSY masses are also included in \textsc{SusHi}. 
The masses, mixing angles, and the effective Yukawa couplings of the Higgs bosons for all scenarios except the hMSSM are calculated with \textsc{FeynHiggs}~\cite{Heinemeyer:1998yj,Heinemeyer:1998np,Degrassi:2002fi,Frank:2006yh,Hahn:2013ria,Bahl:2016brp,Bahl:2017aev}. The branching fractions for the hMSSM scenario are obtained with \textsc{hdecay}~\cite{Djouadi:1997yw,Djouadi:2006bz,Djouadi:2018xqq}, while for all other scenarios the branching fractions are obtained from a combination of \textsc{FeynHiggs}, \textsc{hdecay} and \textsc{PROPHECY4f}~\cite{Bredenstein:2006rh,Bredenstein:2006ha}. The results for the general 2HDM interpretation are obtained using the \ggf cross sections computed with \textsc{SusHi} and the branching fractions with \textsc{2hdmc}~\cite{Eriksson:2009ws}. These calculations are compatible with the results from \textsc{Higlu}~\cite{Spira:1995mt} and \textsc{hdecay} within the uncertainties~\cite{Harlander:2013qxa}. The VBF cross sections are approximated using the SM Higgs boson production cross sections for VBF, which are provided for different masses by the LHC Higgs Cross Section Working Group~\cite{deFlorian:2016spz}, multiplied by $\cos^{2}(\beta-\alpha)$.

\section{Selection and categorization}\label{sec:sel}

At $\sqrts = 13\TeV$, the \ggf cross section for the \hsm is almost one order of magnitude larger than that for VBF production~\cite{deFlorian:2016spz}.
However, the \ggf cross section decreases with \mX while the VBF/\ggf cross section ratio increases, meaning that the VBF production mechanism becomes more important at higher masses.
The main feature distinguishing the two production mechanisms is the presence of associated forward jets for VBF production.
A categorization of events based on both the kinematic properties of associated jets and matrix element techniques is employed to optimize the signal sensitivity.
Events with a VBF topology are selected by requiring the presence of two associated jets with an invariant mass of at least 500\GeV and a $\Delta\eta$ greater than 3.5.

\begin{figure}[htbp]
\centering
\includegraphics[width=0.425\textwidth]{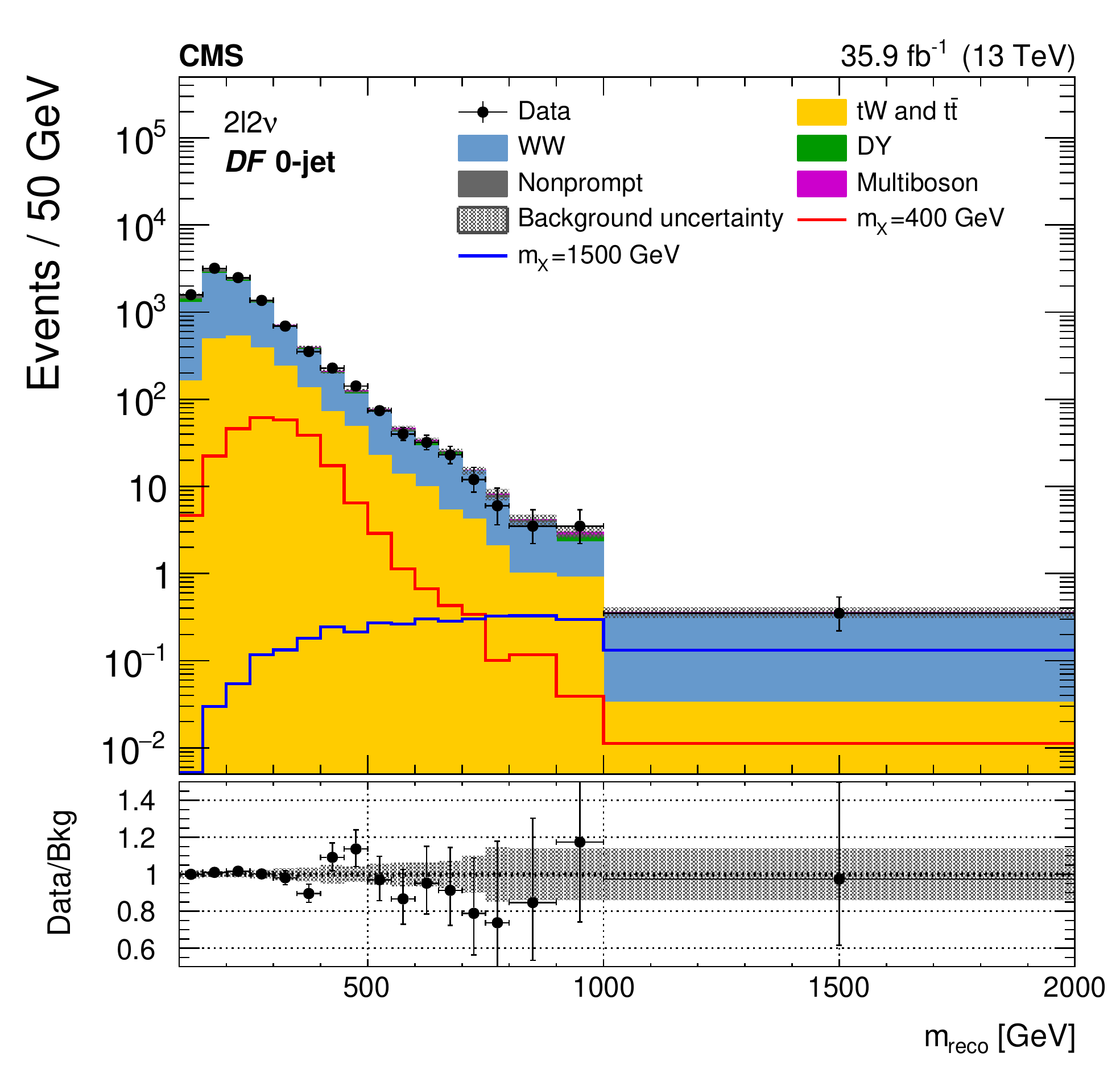}
\includegraphics[width=0.425\textwidth]{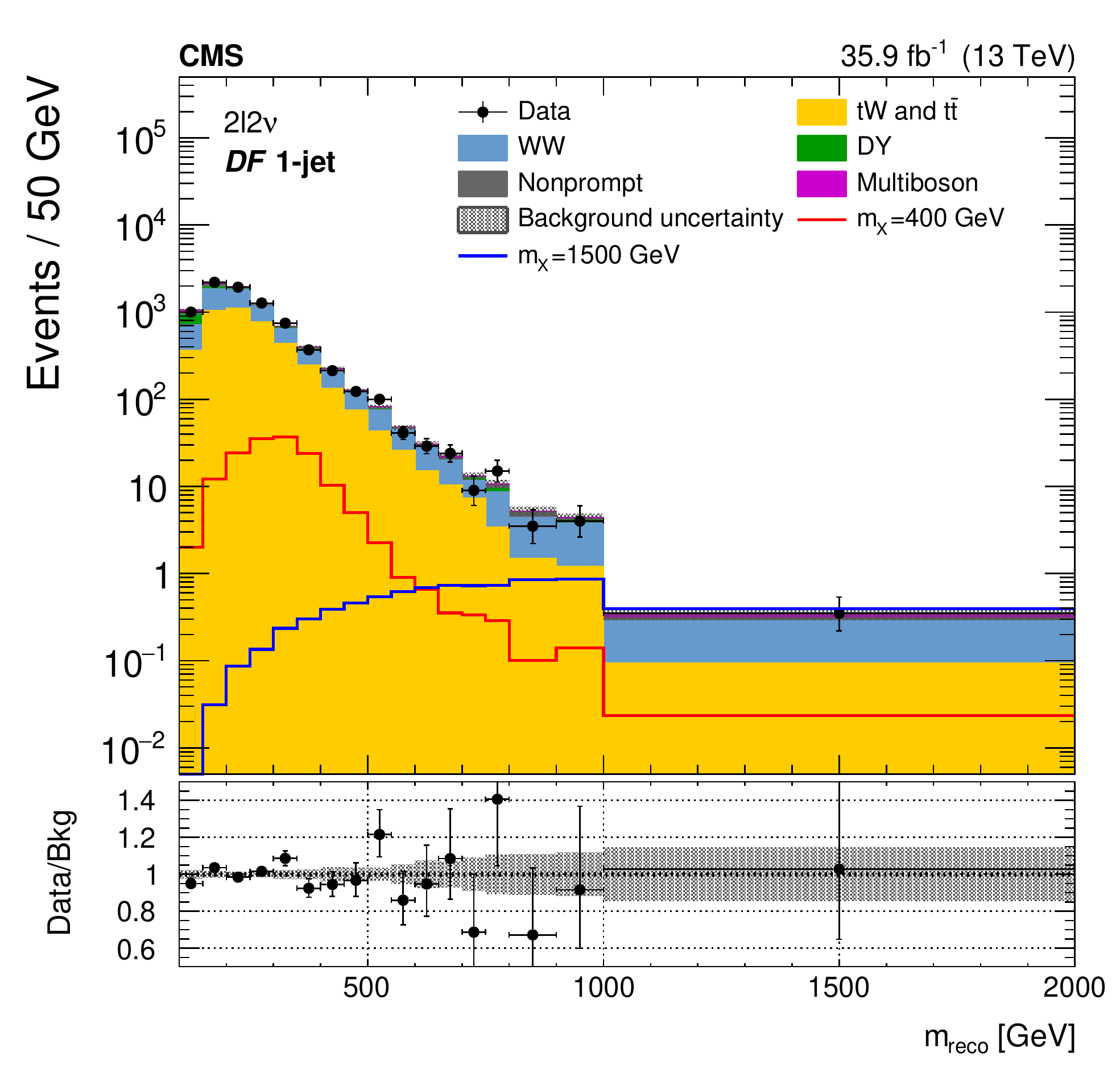}\\
\includegraphics[width=0.425\textwidth]{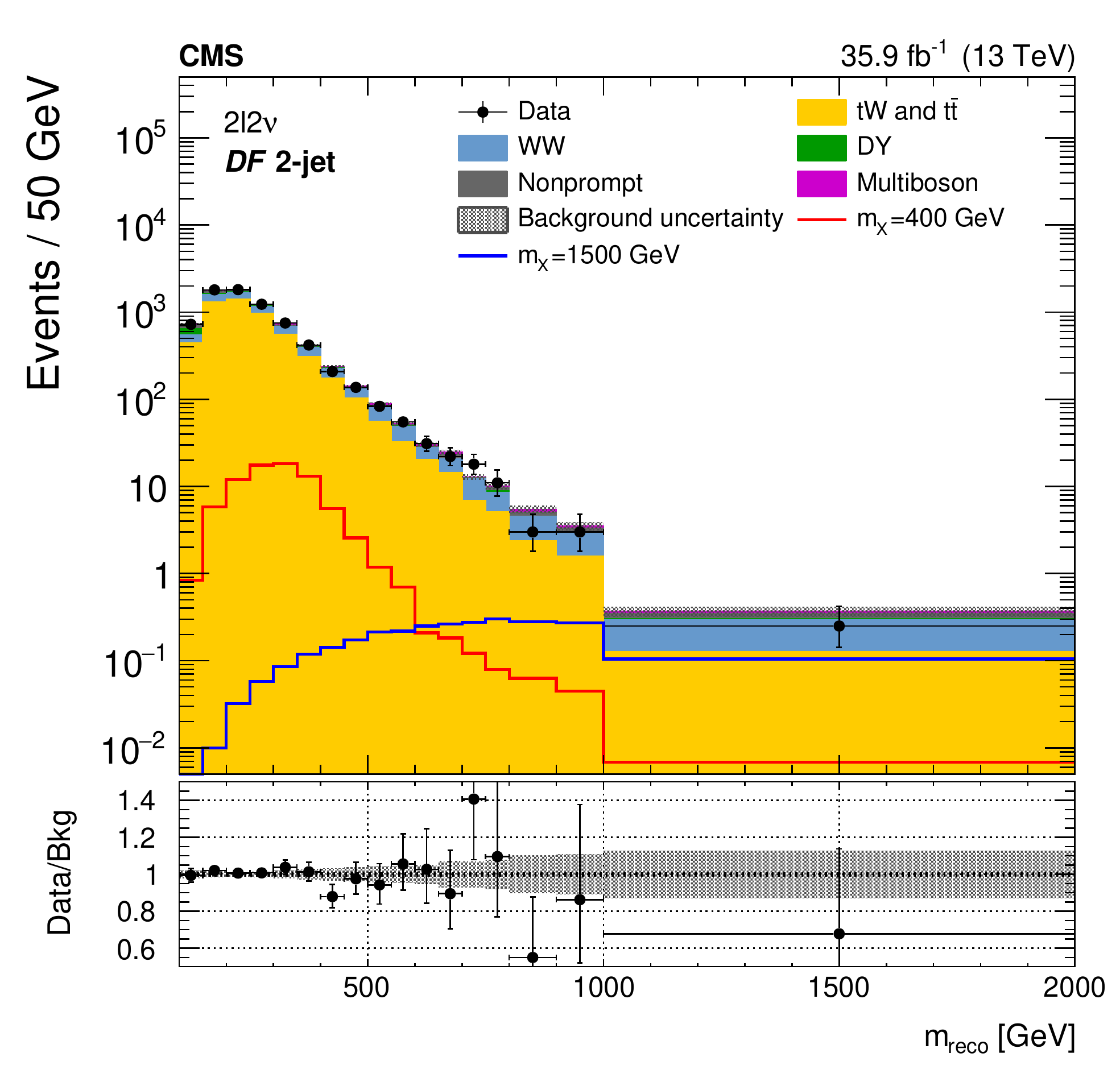}
\includegraphics[width=0.425\textwidth]{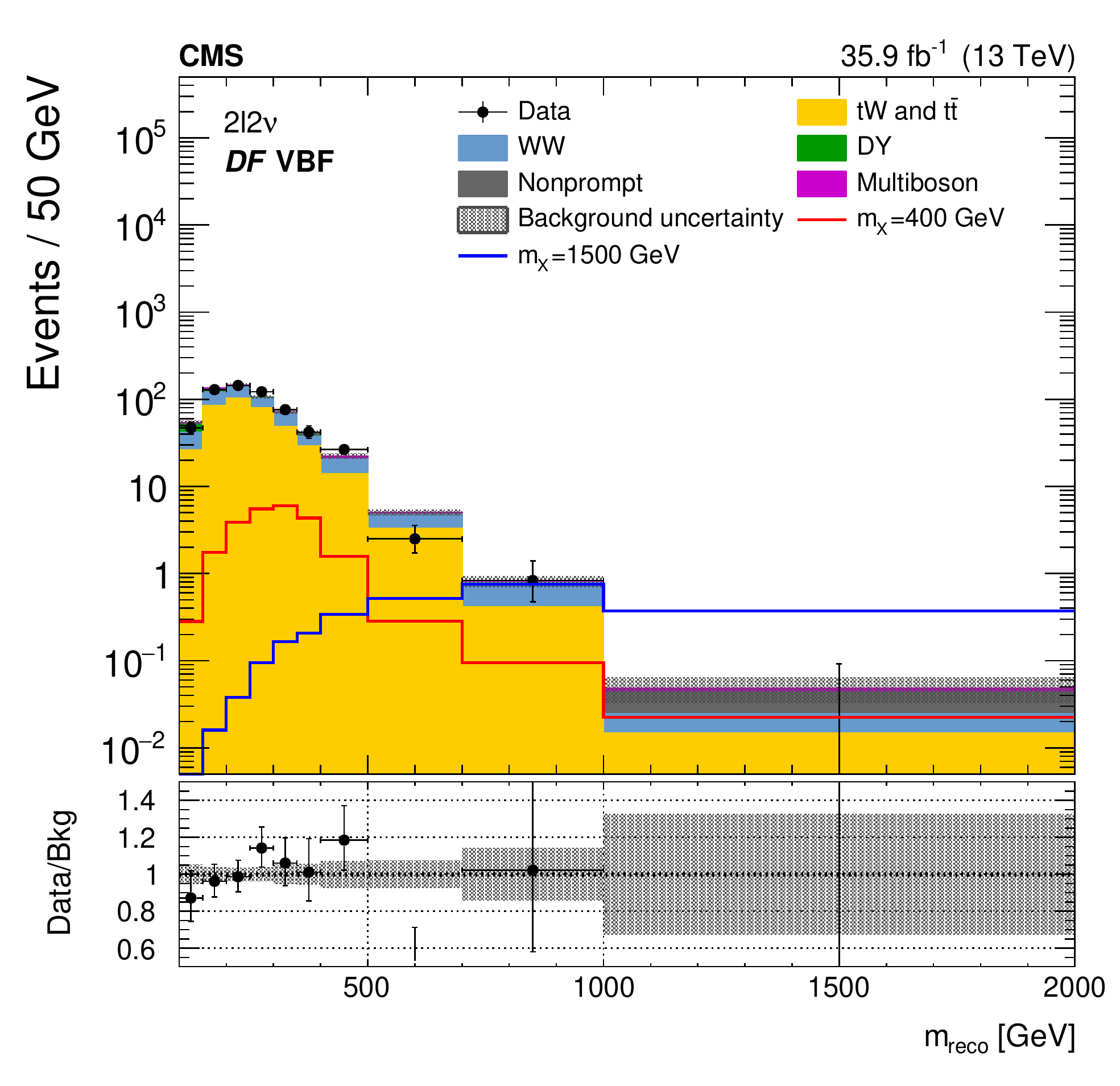}\\
\includegraphics[width=0.425\textwidth]{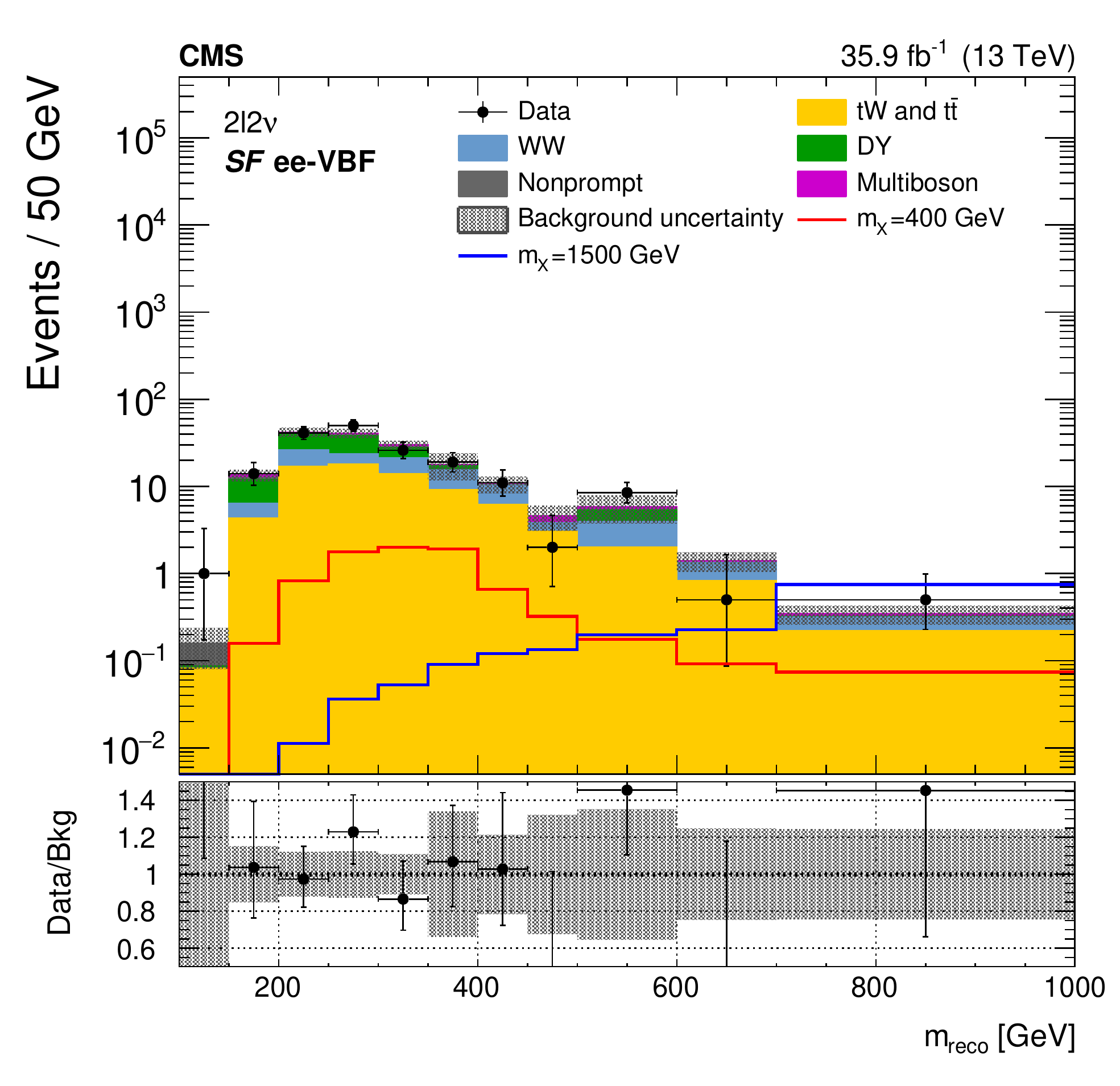}
\includegraphics[width=0.425\textwidth]{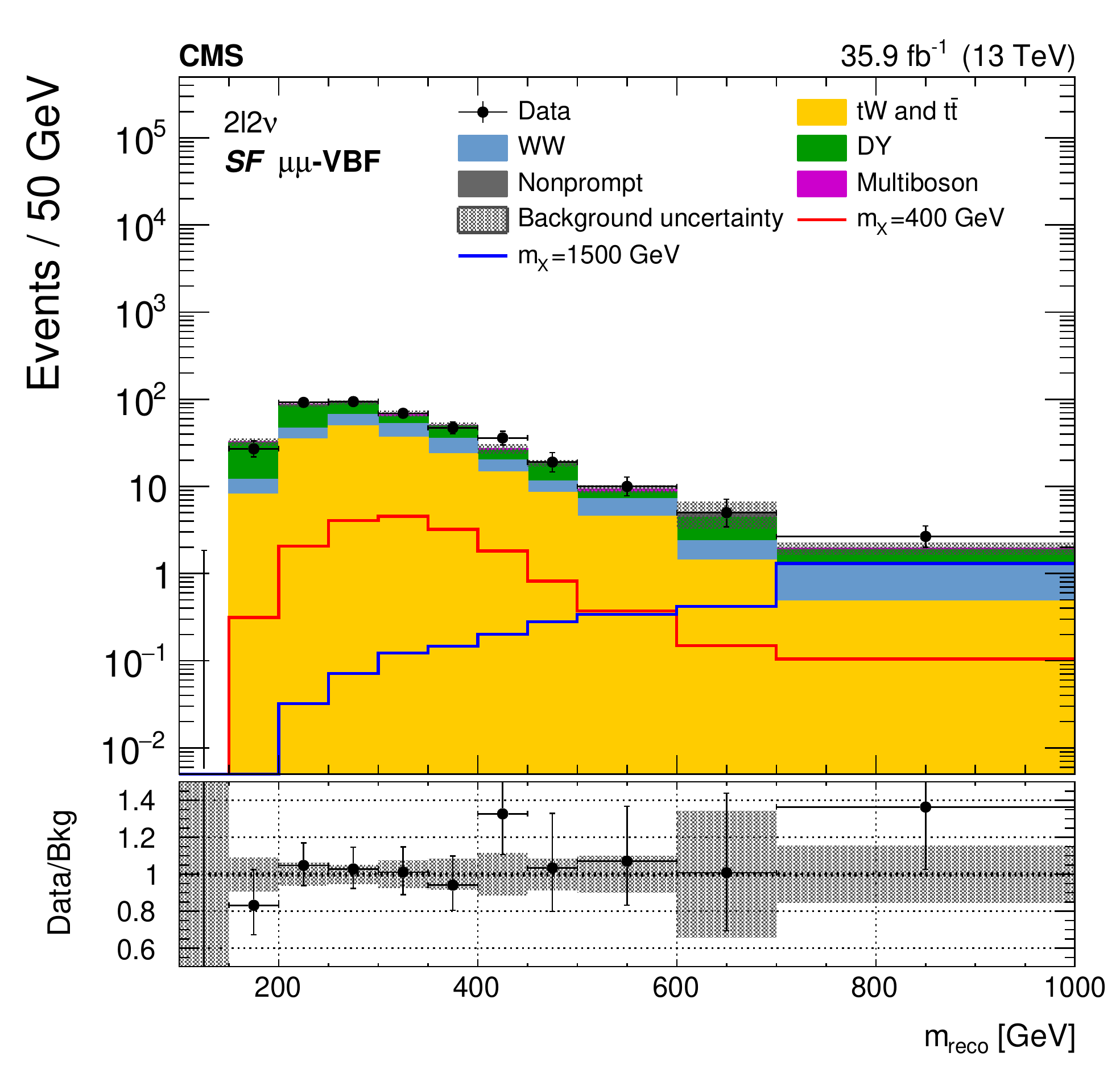}
\caption{
The \mllnn distributions in the \FL different- (upper and middle) and same-flavour (lower) categories, after performing a background-only fit with the dominant background normalizations determined using control regions.
The points represent the data and the stacked histograms the expected backgrounds.
Also shown are the sum of the expected \ggf- and VBF-produced signals for $\mX = 400$ and 1500\GeV, normalized to the SM cross sections, and without considering interference effects.
The hatched area shows the combined statistical and systematic uncertainties in the background estimation.
Lower panels show the ratio of data to expected background.
Larger bin widths are used at higher \mllnn; the bin widths are indicated by the horizontal error bars.
}
    \label{fig:mti_sig}
\end{figure}

\subsection{ \texorpdfstring{\PX \ra \FL}{2l2nu} }

\tolerance=400{
The \FL analysis selects two oppositely charged leptons in the same- and different-flavour final states.
To suppress the background from nonprompt leptons arising from \Wjets production, both leptons must be well identified and isolated.
Events are categorized according to the lepton flavour composition and the number of AK4 jets with $\pt > 30\GeV$.
To suppress the top quark background, events are required to have no \bq-tagged AK4 jets with $\pt > 20\GeV$.
The final discriminating variable is the reconstructable mass $\mllnn = \sqrt{\smash[b]{ (p_{\ell\ell} + \ptmiss)^2 - (\vec{p}_{\ell\ell} + \ptvecmiss)^2 }}$, where $(p_{\ell\ell}, \vec{p}_{\ell\ell})$ is the dilepton four-momentum.
This variable is chosen for its effectiveness in discriminating between signal and background, and between different signal mass hypotheses.
}

\subsubsection{Different-flavour final state}

For the different-flavour \emu channel, one of the two leptons is required to have $\pt > 25\GeV$ and the other is required to have $\pt > 20\GeV$.
To suppress background processes with three or more leptons in the final state, such as \ZZ, \WZ, or triboson production, events with an additional identified and isolated lepton with $\pt > 10\GeV$ are rejected.
The dilepton invariant mass \mll is required to be higher than 50\GeV to reduce the \hsm contamination.
Due to the presence of neutrinos in the final state of interest, only events with $\ptmiss > 20\GeV$ are considered.
The DY \ra \tautau background is suppressed by requiring that the dilepton transverse momentum \ptll is above 30\GeV and the \PX transverse mass \mthll is above 60\GeV, where $\mthll  = \sqrt{\smash[b]{ 2 \ptll \ptmiss(1 - \cos \dpll) }}$ and \dpll is the azimuthal angle between \ptvecmiss and \ptvecll.
Finally, motivated by the high-mass of the signals under investigation, the condition $\mllnn > 100\GeV$ must be satisfied.

In this channel four exclusive jet categories are defined: a zero-jet, one-jet, two-jet and VBF category.
The last category requires the presence of exactly two jets which satisfy the VBF selection criteria.
Dijet events failing these criteria enter the two-jet category.
Figure~\ref{fig:mti_sig} displays the \mllnn distributions for events passing the \FL different-flavour selection in the four exclusive jet categories.

\subsubsection{Same-flavour final state}

For the same-flavour \elel and \mumu channels, both leptons are required to have $\pt > 20\GeV$.
Events with an additional identified and isolated lepton with $\pt > 10\GeV$ are rejected.
The background rejection requirements described for the \emu channel are also applied in these channels.
To suppress the large DY \ra \elel and DY \ra \mumu backgrounds only those events with two jets satisfying the VBF selection criteria are considered.
For the further reduction of this background, the \mll and \ptmiss requirements are raised to 120 and 50\GeV, respectively.
Figure~\ref{fig:mti_sig} displays the \mllnn distributions for events passing the \FL same-flavour selection.

\subsection{\texorpdfstring{\PX \ra \SL}{lnu2q}}

In the \SL analysis, the \W \ra \LN candidates are reconstructed by combining the \ptmiss and a lepton which has $\pt > 30\GeV$ and $\abs{\eta} < 2.1$ (2.4) for electrons (muons).
Those events containing additional electrons (muons) with $\pt > 15$ (10)\GeV passing loose identification requirements are rejected.
The \ptmiss is considered as an estimate of the neutrino \pt with the longitudinal component \pz of the neutrino momentum estimated by imposing a \W boson mass constraint to the \LN system and solving the corresponding quadratic equation.
The solution with the smallest magnitude of neutrino \pz is chosen. When a real solution is not found, only the real part is considered.
The \W \ra \wqqb candidates are reconstructed as either high-\pt merged jets or as resolved low-\pt jet pairs.
A \W boson mass window selection is applied to suppress the \Wjets background.
If an additional AK4 jet with $\pt > 20\GeV$ which is \bq-tagged is present, then the event is rejected to suppress the top quark background.
The \W \ra \LN and \W \ra \wqqb decay candidates are combined into \WW resonance candidates.
The final discriminating variable is the invariant mass of the \WW system, \mww.

Events are categorized based on the tagging of VBF and \ggf production mechanisms.
A VBF category is defined by requiring two additional AK4 jets satisfying the VBF selection criteria.
Those events failing the VBF selection are considered for the \ggf category.
The tagging of \ggf candidates is achieved using a kinematic discriminant based on the angular distributions
of the \PX candidate decay products. This is implemented with MELA which uses \textsc{JHUGen} and \MCFM
matrix elements to calculate probabilities for an event to come from either signal or background, respectively.
A \WW resonance candidate is considered \ggf-tagged if the kinematic discriminant is greater than 0.5.
Those events with \WW resonance candidates failing this requirement enter the untagged category, resulting in three production mechanism categories.

\subsubsection{Boosted final state}

\begin{figure}[htbp]
\centering
\includegraphics[width=0.425\textwidth]{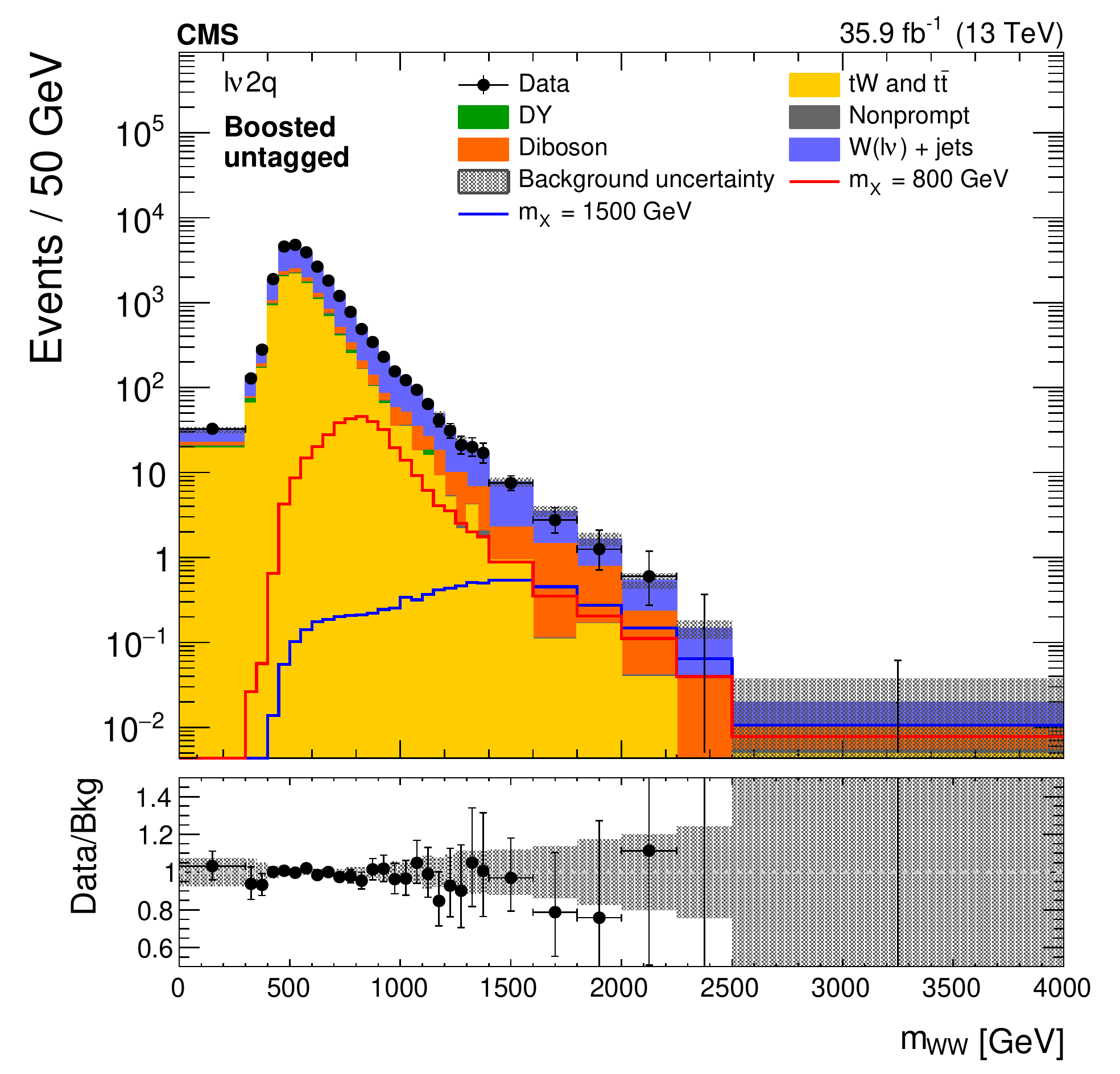}
\includegraphics[width=0.425\textwidth]{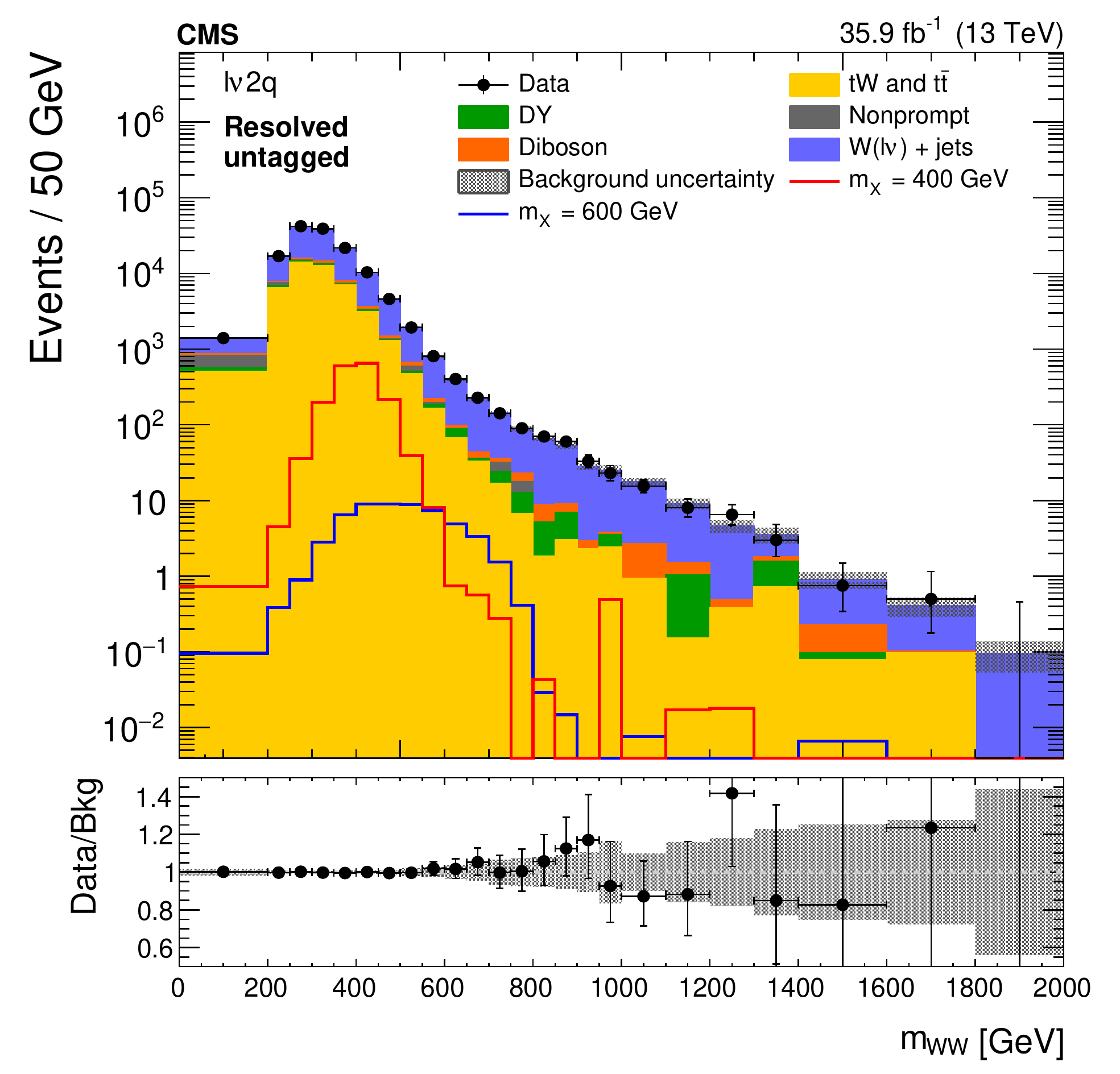} \\
\includegraphics[width=0.425\textwidth]{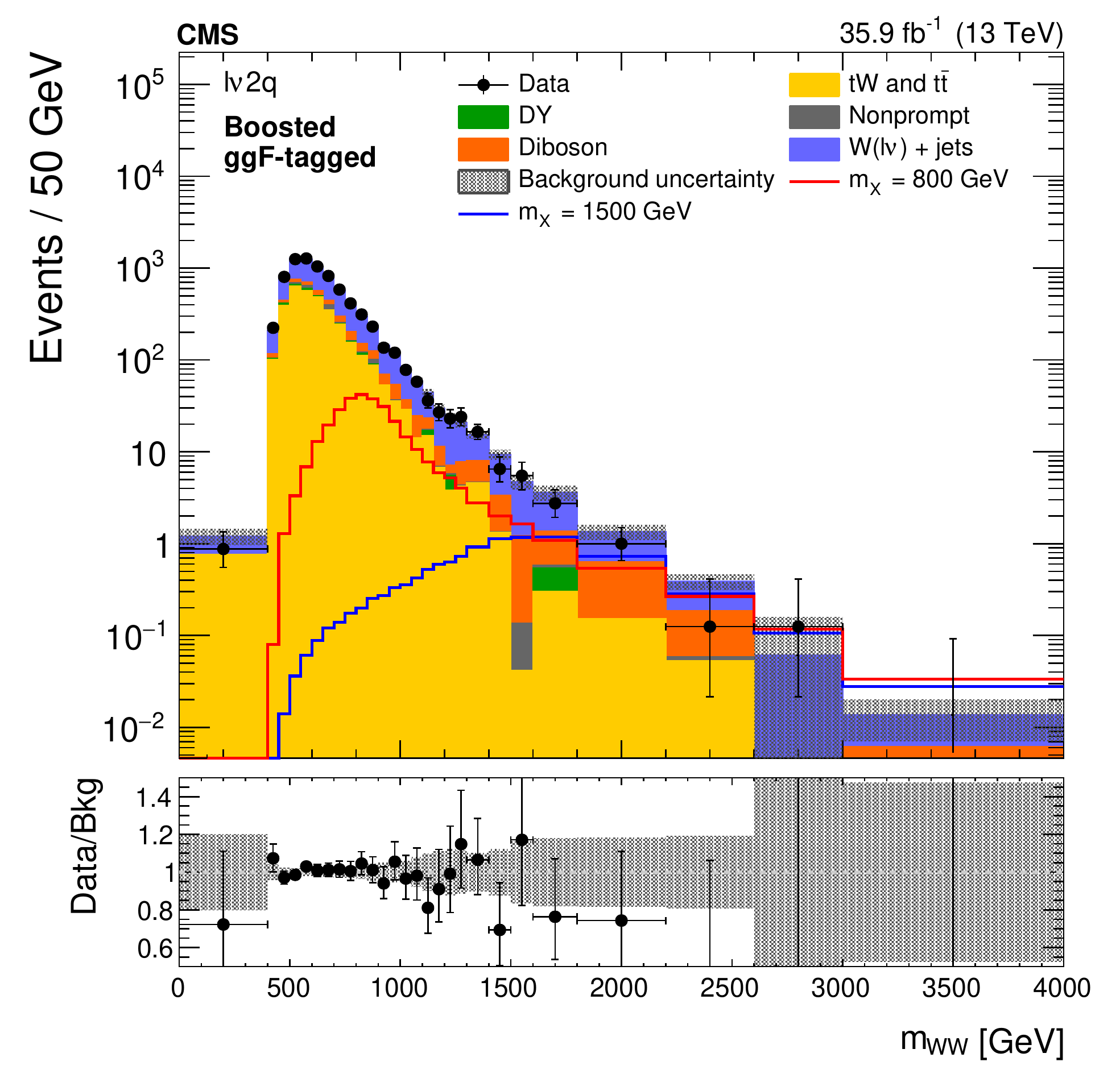}
\includegraphics[width=0.425\textwidth]{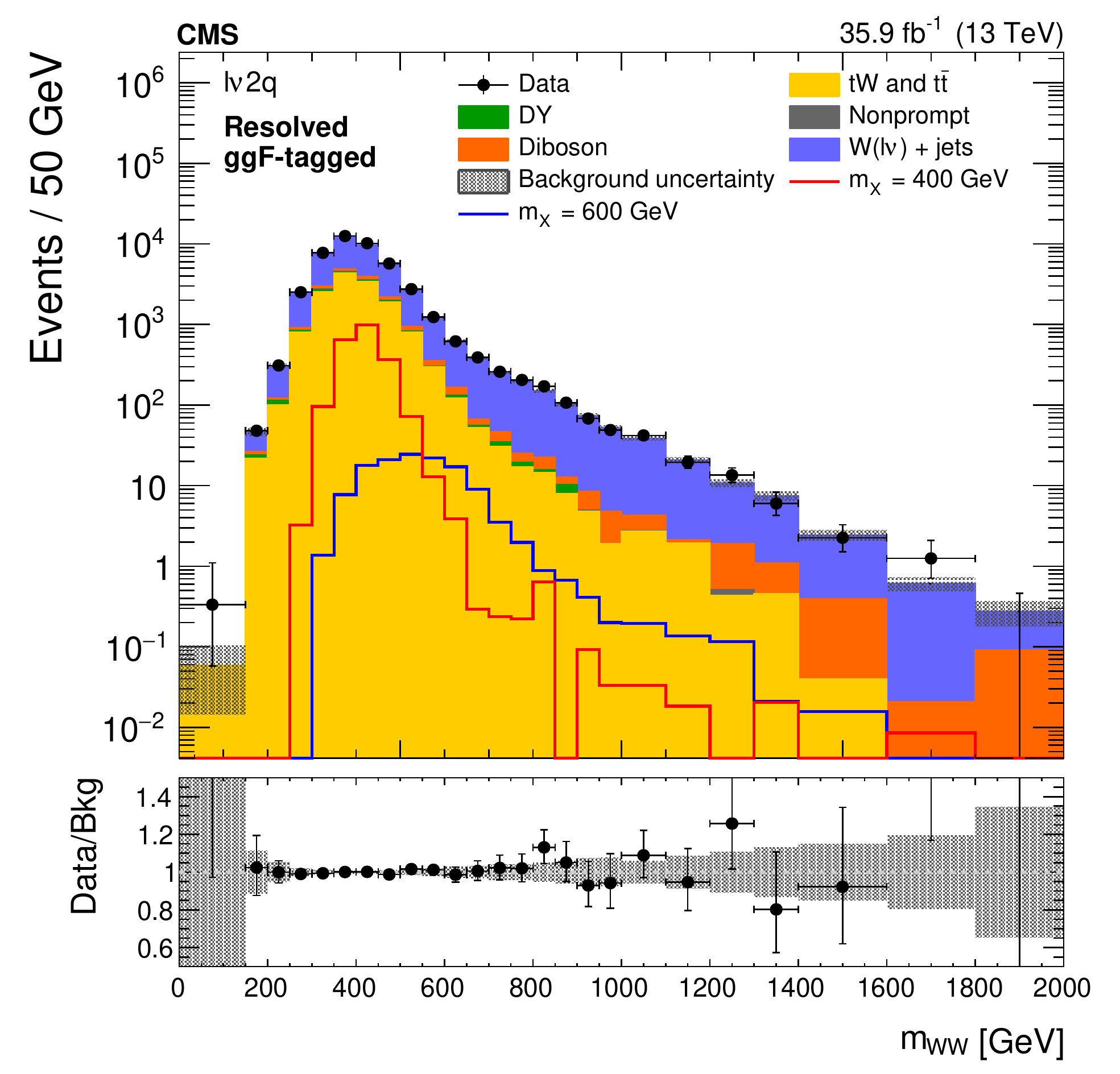} \\
\includegraphics[width=0.425\textwidth]{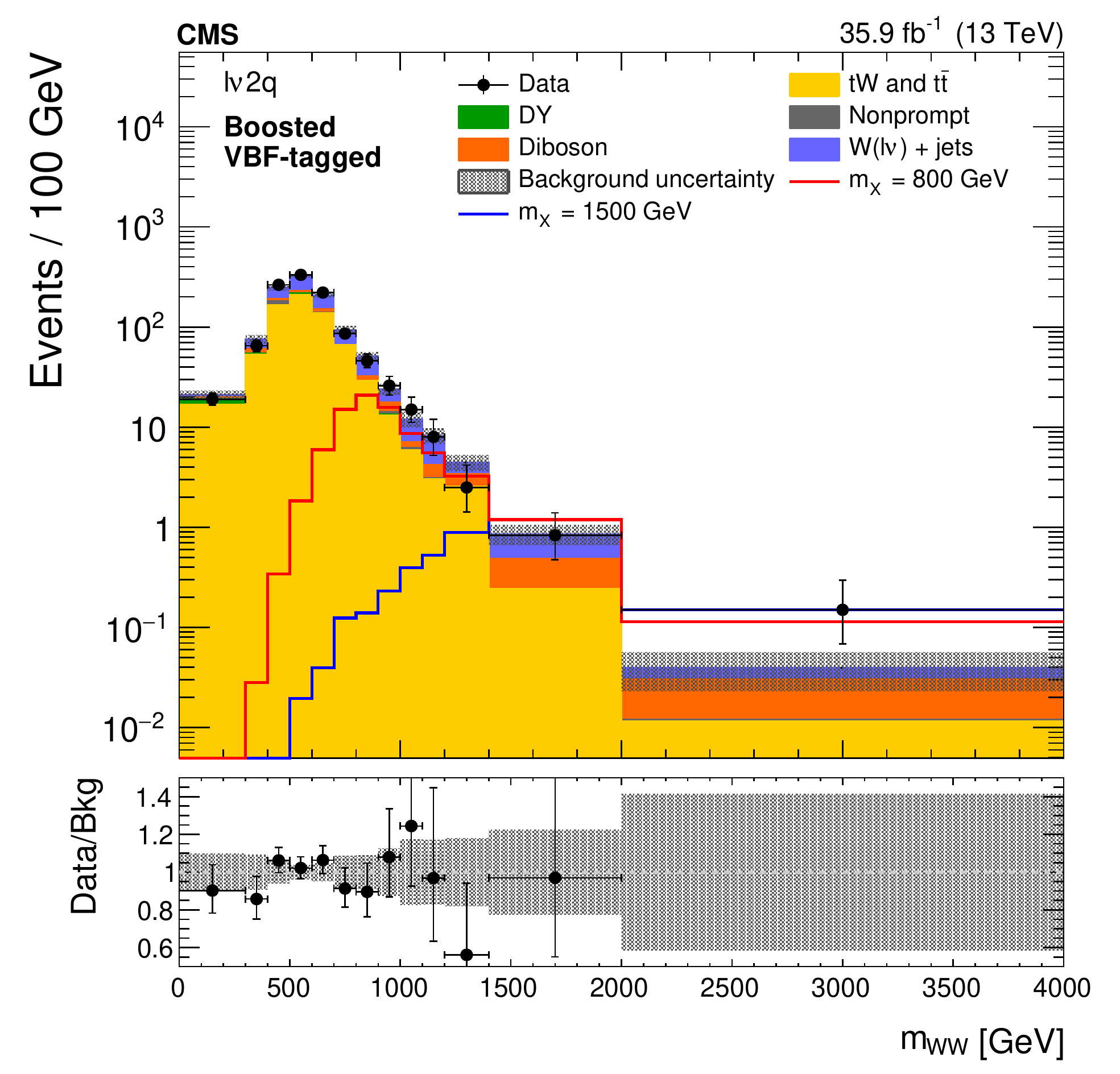}
\includegraphics[width=0.425\textwidth]{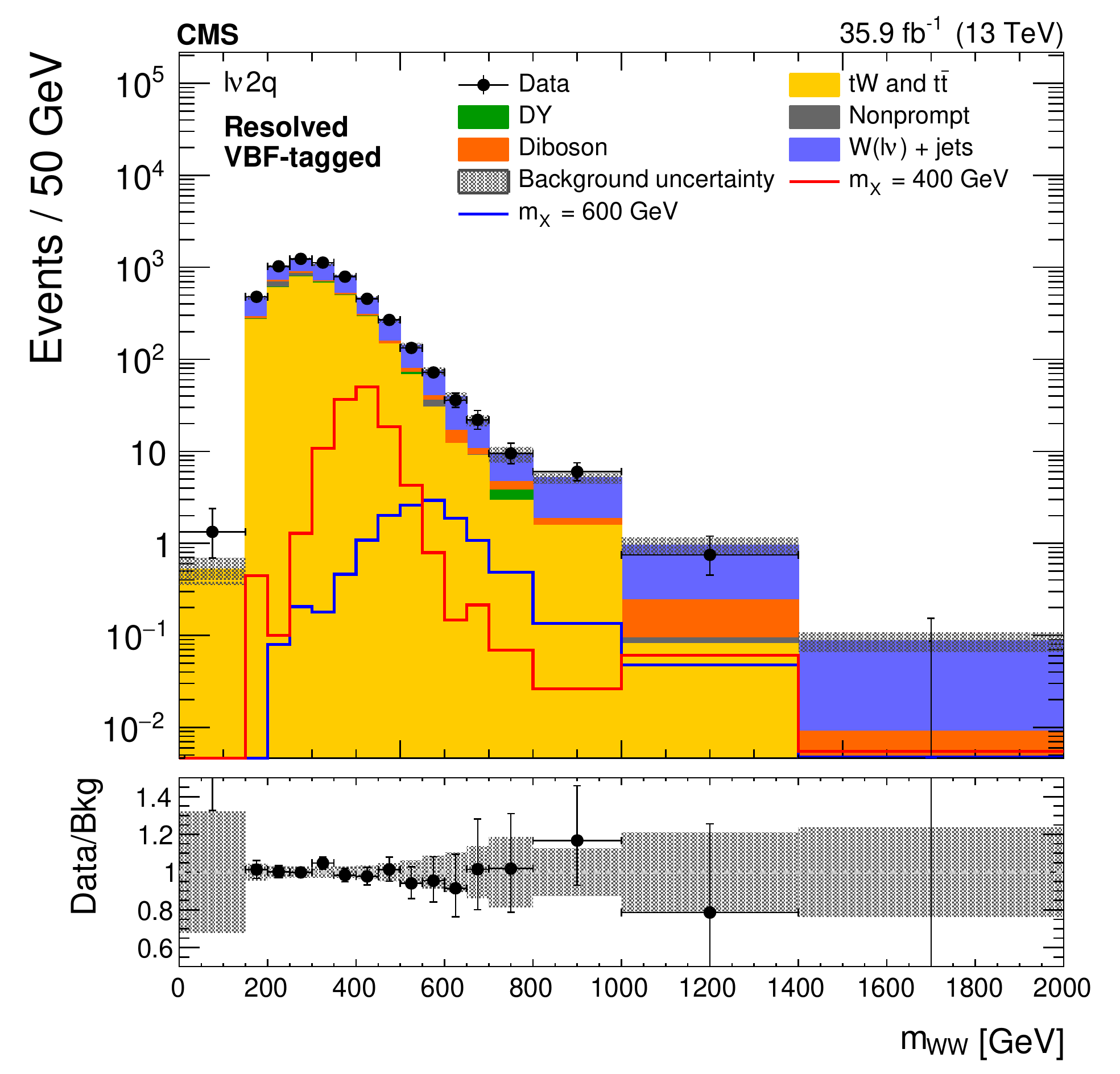} \\
\caption{
The \mww distributions in the \SL boosted (left) and resolved (right) categories, after performing a background-only fit with the dominant background normalizations determined using control regions. Electron and muon channels are combined.
The points represent the data and the stacked histograms the expected backgrounds.
Also shown are the sum of the expected \ggf- and VBF-produced signals for $\mX = 800$ and 1500\GeV (left), and  $\mX = 400$ and 600\GeV (right), normalized to the SM cross sections, and without considering interference effects.
The hatched area shows the combined statistical and systematic uncertainties in the background estimation.
Lower panels show the ratio of data to expected background.
Larger bin widths are used at higher \mww; the bin widths are indicated by the horizontal error bars.
}
\label{fig:HMass_lnuqq}
\end{figure}

For the boosted final state, an AK8 jet with \mj in the mass window $65 < \mj < 105\GeV$ is required.
To suppress the background from nonprompt leptons in QCD multijet events, only events with $\ptmiss > 40\GeV$ are considered.
For heavy-resonance decays the \pt of the \W candidates are expected to be roughly half of the resonance mass.
Therefore both the leptonic and hadronic \W candidates must satisfy the condition $\ptw/\mww > 0.4$.
Finally, to identify boosted \W candidates (boosted \W tagging) the $N$-subjettiness ratio \tauto is required to be $<$0.4.
The \mww distributions for events passing the \SL boosted selection in the three production categories are shown in Fig.~\ref{fig:HMass_lnuqq}.

\subsubsection{Resolved final state}

For events that do not contain a boosted \W-tagged jet with $\mj > 40\GeV$, a resolved hadronic \W boson decay reconstruction is attempted using two AK4 jets with $\pt > 30\GeV$ and $\abs{\eta} < 2.4$.
In events with greater than two jets the selection of the dijet pair is performed by means of a kinematic fit~\cite{Abbott:1998dc}.
For each dijet pair the kinematic fit algorithm constrains the jet four-momenta, assuming the dijet invariant mass is that of the \W boson, and assigns a \chisq according to the goodness of the fit.
The dijet pair with the smallest \chisq is chosen as the hadronic \W candidate.
The invariant mass of the dijet system must be in the mass window $65 < \mjj < 105\GeV$.
To suppress the background from nonprompt leptons in QCD multijet events, it is required that $\ptmiss > 30\GeV$ and that the leptonic \W candidate transverse mass \mtw is above 50\GeV, where $\mtw  = \sqrt{\smash[b]{ 2 \ptl \ptmiss(1 - \cos \dpl) }}$ and \dpl is the azimuthal angle between \ptvecmiss and the lepton transverse momentum \ptvecl. The leptonic and hadronic \W candidates must also satisfy the condition $\ptw/\mww > 0.35$.
Further reduction in the QCD multijet background is achieved by requiring that the \PX transverse mass \mthljj is above 60\GeV, where $\mthljj = \sqrt{\smash[b]{ 2 \ptljj \ptmiss(1 - \cos \dpljj) }}$ and \dpljj is the azimuthal angle between \ptvecmiss and the transverse momentum of the lepton plus jets system \ptvecljj.
The \mww distributions for events passing the \SL resolved selection in the three production categories are shown in Fig.~\ref{fig:HMass_lnuqq}.

\section{Background estimation}\label{sec:bkg}

The dominant backgrounds are modeled via simulation that has been reweighted to account for known discrepancies between data and simulated events.
Corrections associated with the description in simulation of the trigger efficiencies, as well as the efficiency for electron and muon reconstruction, identification, and isolation, are extracted from events with leptonic \PZ boson decays using a ``tag-and-probe'' technique~\cite{CMS:2011aa}. The \bq tagging efficiency is measured using data samples enriched in \bq quark jets and corrections for simulation derived~\cite{Chatrchyan:2012jua}. For the \SL boosted category, corrections are applied to the \W tagging efficiency and the \mj scale and resolution of \W-tagged jets. These corrections have been measured in an almost pure sample of semileptonic \ttbar events, where boosted \W bosons produced in the top quark decays are separated from the combinatorial \ttbar background by means of a simultaneous fit to \mj~\cite{Khachatryan:2014vla}.
For the normalization of the major backgrounds data driven estimates using control regions are employed.

\subsection{\texorpdfstring{\PX \ra \FL}{2l2nu}}

The main background processes contributing to the \FL final state are from nonresonant \WW and top quark production.
The nonresonant \WW background populates the entire phase space in \mllnn while the high-mass signal contribution is concentrated at high values of this variable.
Therefore, this background is estimated directly in the final fit to the data by allowing the \WW normalization to float freely and independently in each category.

\begin{figure}[htbp]
\centering
\includegraphics[width=0.44\textwidth]{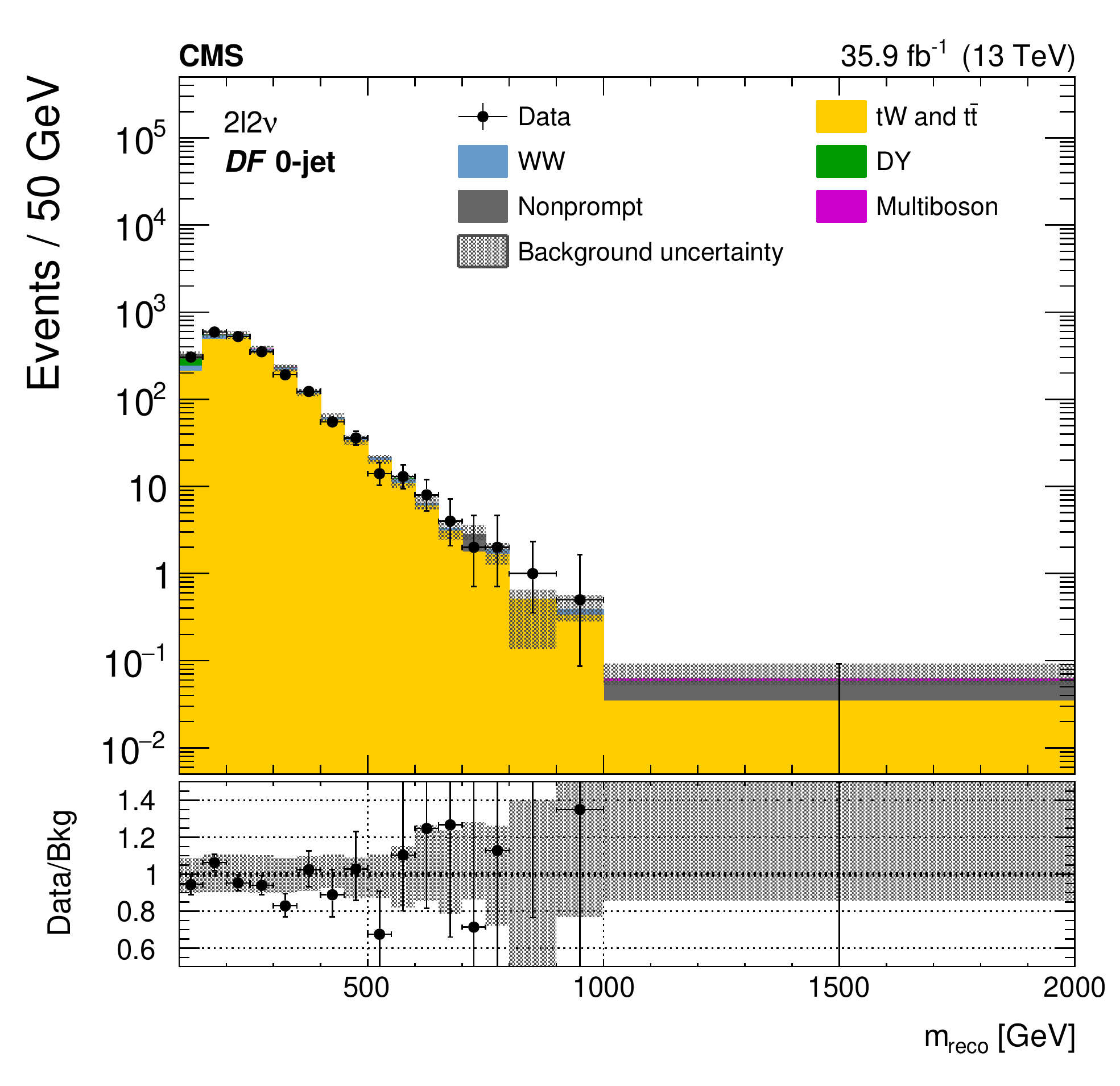}
\includegraphics[width=0.44\textwidth]{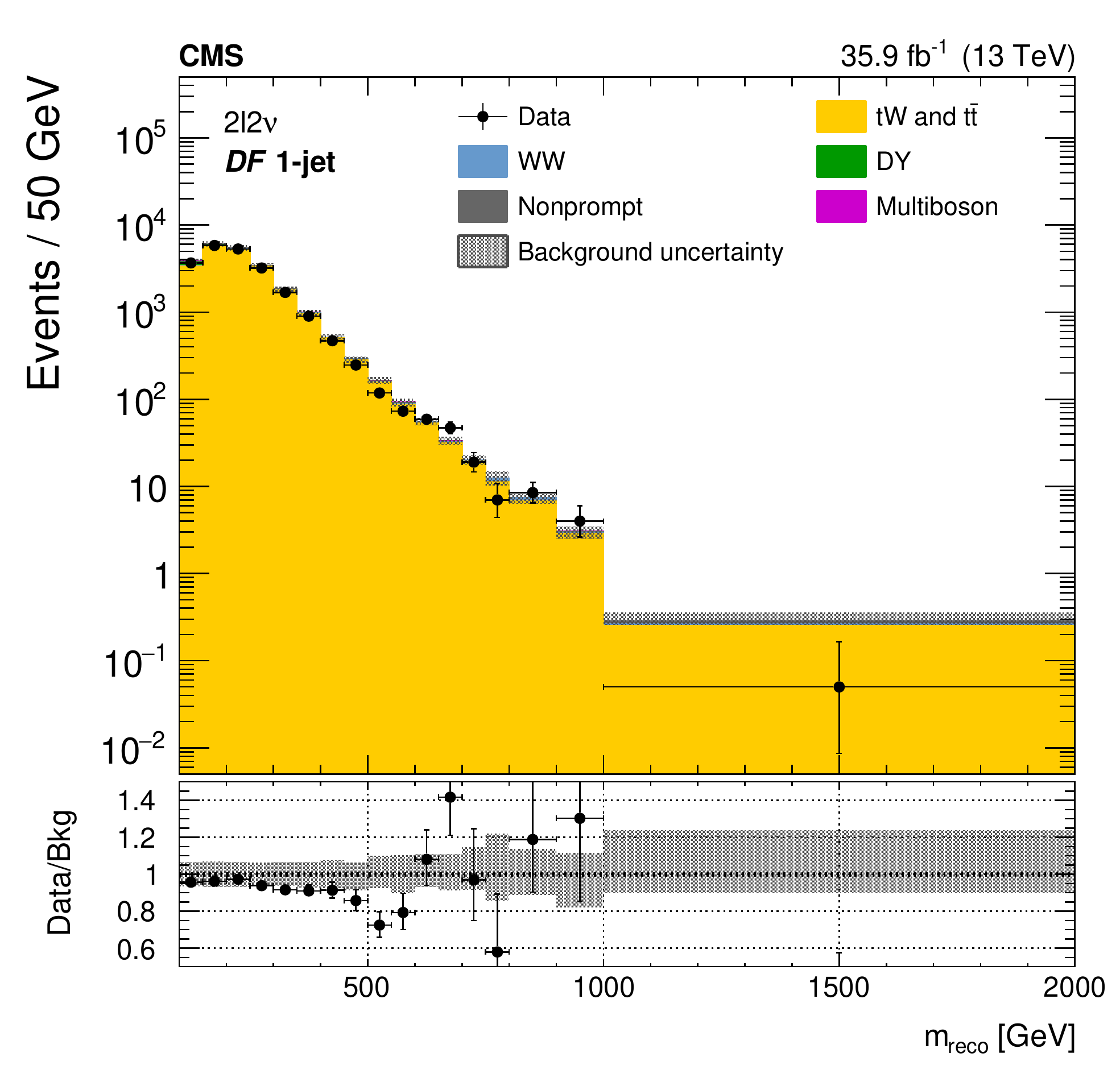}\\
\includegraphics[width=0.44\textwidth]{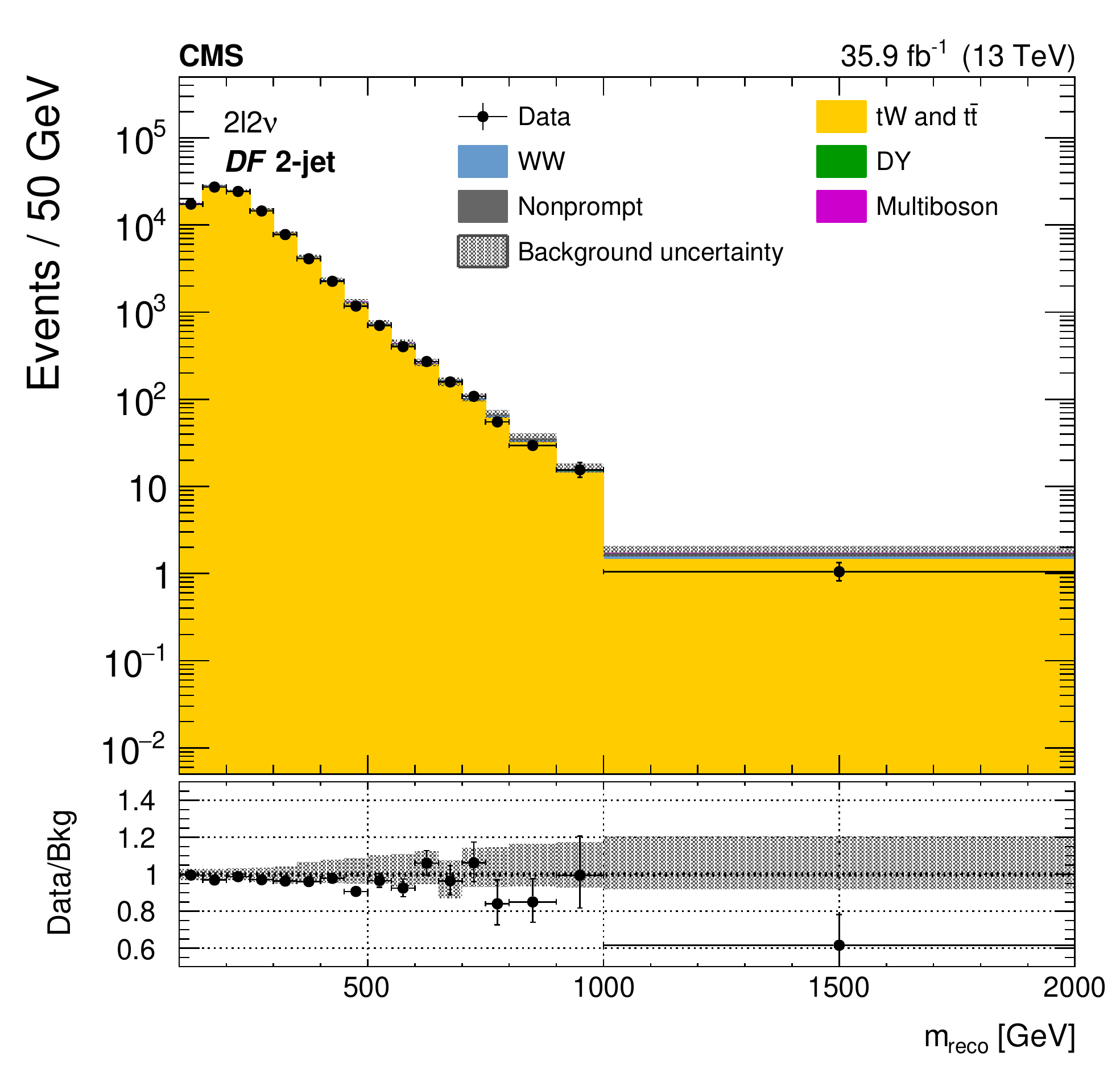}
\includegraphics[width=0.44\textwidth]{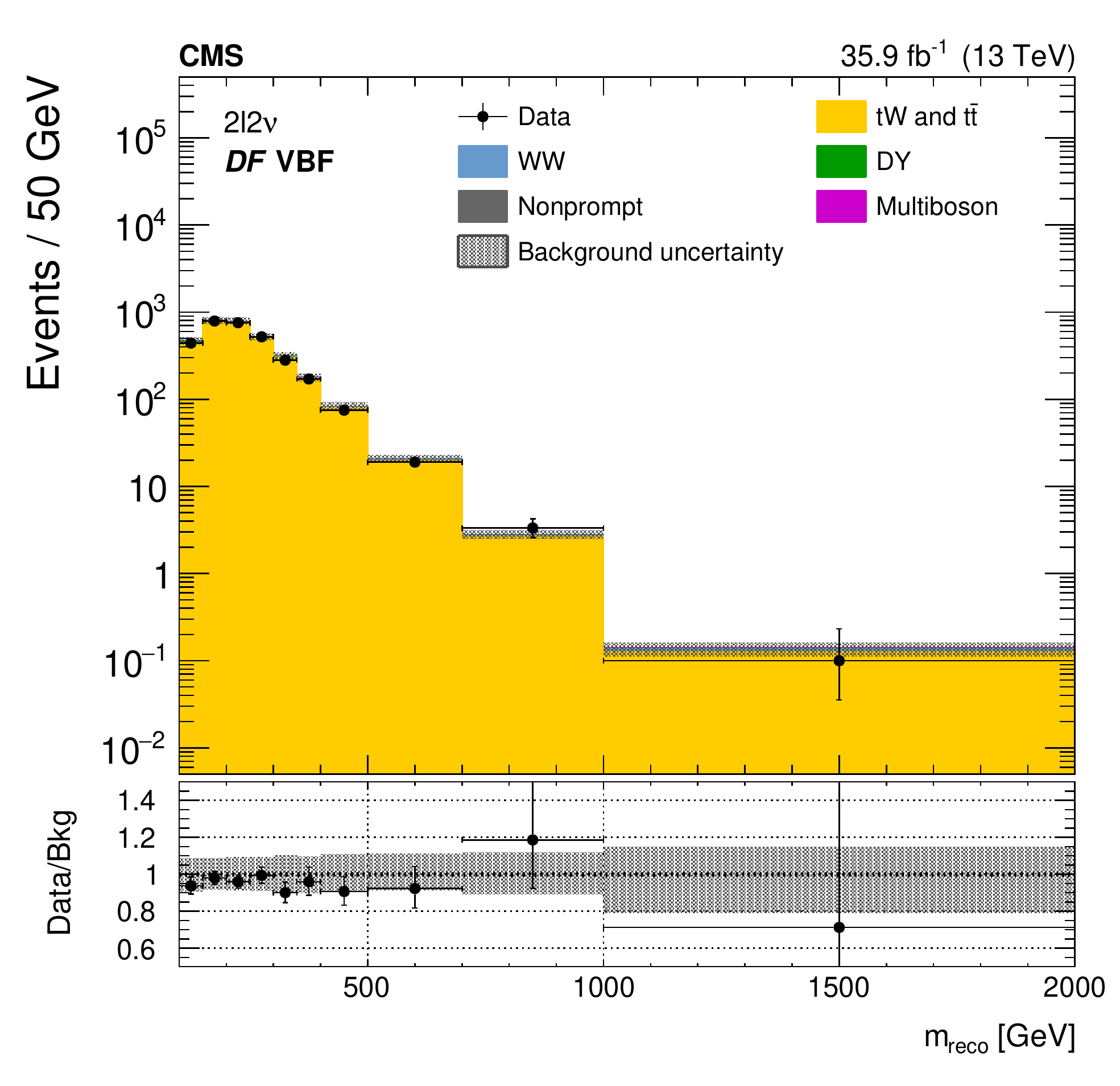}\\
\includegraphics[width=0.44\textwidth]{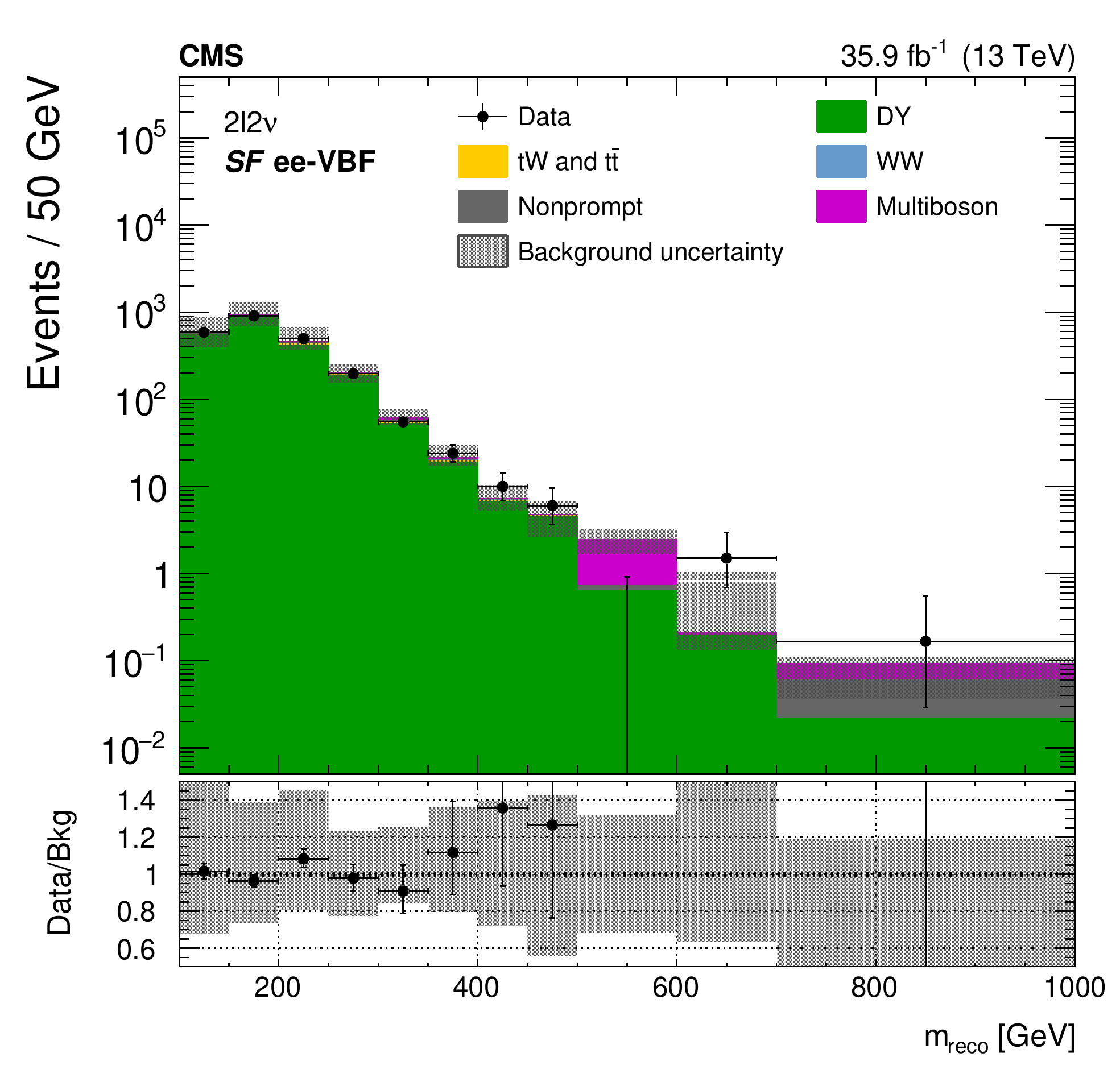}
\includegraphics[width=0.44\textwidth]{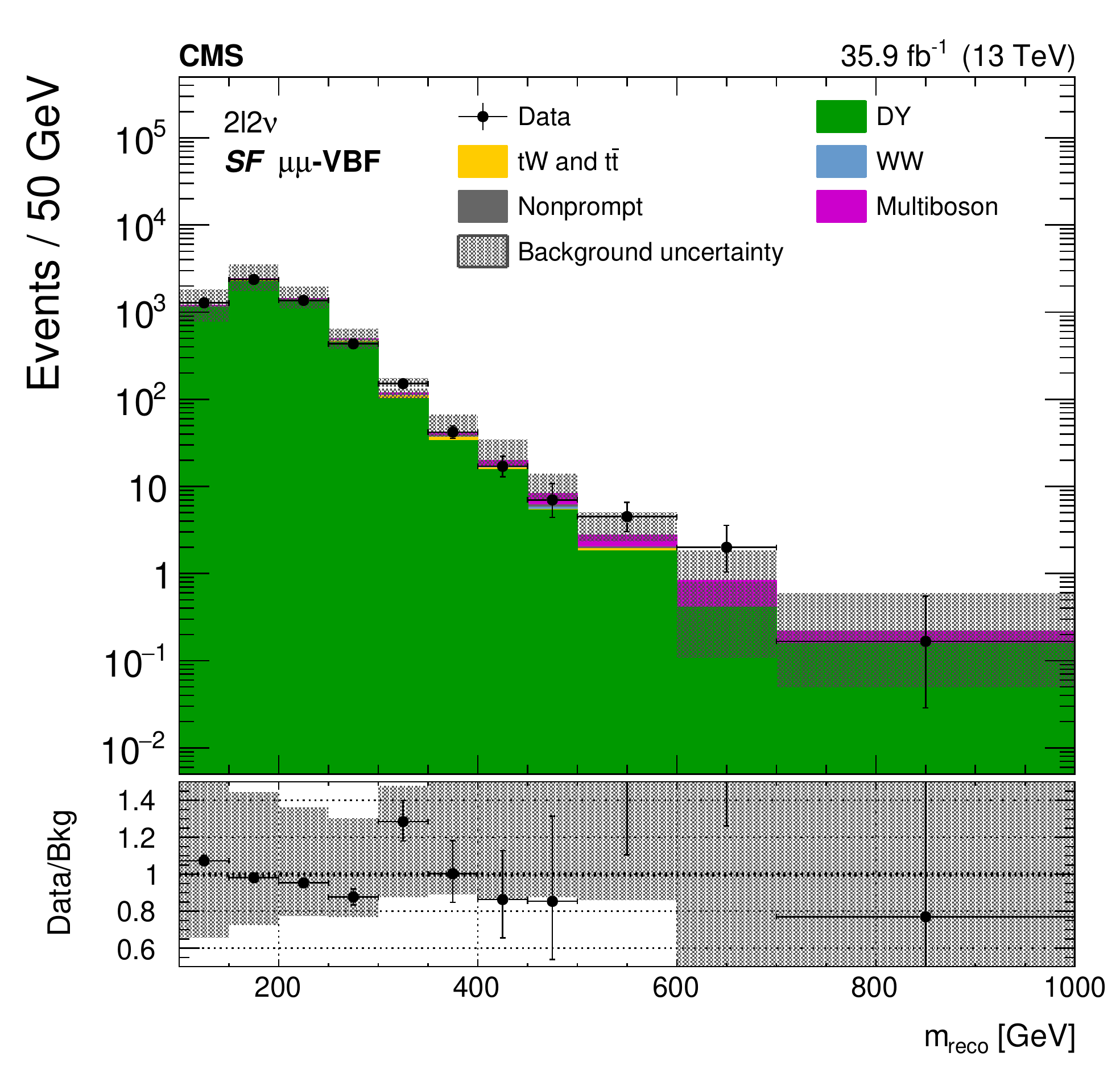}
\caption{The \mllnn distributions in the top quark control regions of the \FL different-flavour categories (upper and middle) and the DY control regions of the \FL same-flavour categories (lower).
The points represent the data and the stacked histograms show the expected backgrounds.
The hatched area shows the combined statistical and systematic uncertainties in the background estimation.
Lower panels show the ratio of data to expected background.
Larger bin widths are used at higher \mllnn; the bin widths are indicated by the horizontal error bars.
}
\label{fig:mti_top}
\end{figure}

The estimation of the top quark background is performed using a top quark enriched data control region, defined by inverting the \bq jet veto requirement.
It is used to constrain the top quark background normalization which is allowed to float freely in the final fit to the data.
The estimation is performed separately for each of the different- and same-flavour categories.
The \mllnn distributions in the top quark control regions of each of the different-flavour categories are shown in Fig.~\ref{fig:mti_top}.
The expected backgrounds before fitting the data are shown, good agreement between the top quark background predictions and the data is observed.

The DY process is a significant source of background in the same-flavour categories.
A subleading source of background in the different-flavour categories comes from DY \ra \tautau, where each \Pgt decays leptonically.
In the final fit to the data, the DY normalization is also allowed to float freely and independently in each category, and is constrained using control regions which are defined using modified signal region selections.
For the different-flavour channel, a DY control region is defined for each jet category by inverting the signal region \mthll selection, requiring $\mthll < 60\GeV$.
The invariant mass of the two leptons is restricted to the interval between 50 and 80\GeV to reduce contributions from nonprompt leptons and from top quark processes.
For the same-flavour channels, the control regions are defined by changing the signal region \mll selection to require $70 < \mll < 120\GeV$.
Discrepancies are observed between the \ptmiss distributions in data and simulation for the same-flavour control regions.
A linear \ptmiss correction is derived for the simulation by fitting the ratio between data, with minor background subtracted, and the DY prediction.
The \mllnn distributions in the DY control regions of each of the same-flavour categories are shown in Fig. \ref{fig:mti_top}.
The expected backgrounds before fitting the data are shown, good agreement between the DY background predictions and the data is observed.

The instrumental background arising from nonprompt leptons in \Wjets production is estimated to be between 2 and 8\% of the total background.
An estimate is done in a control region that uses looser lepton identification criteria with relaxed isolation requirements.
The probability for a jet that satisfies the loose lepton requirements to also satisfy the standard selection is determined using dijet events.
Similarly, the efficiency for a prompt lepton that satisfies the loose lepton
identification requirements to also satisfy the standard selection is determined using DY events.
These efficiencies are then used to weight the data events with the probability for the event to contain a nonprompt lepton and the relative probability for the candidates in this event to also satisfy the standard selection.
Other subleading backgrounds, such as \WZ, \ZZ, and triboson production, are estimated from simulation.

\subsection{ \texorpdfstring{\PX \ra \SL}{lnu2q}}

The main backgrounds for the \SL analysis are from \Wjets and top quark production, with subdominant contributions from diboson, DY, and QCD multijet production.

The majority of the events passing the \SL selection come from \Wjets and top quark production.
An estimate of the \Wjets and top quark background normalizations using two control regions in data is employed.
A top quark enriched data control region is defined reversing the \bq jet veto, by requiring events with an additional jet which is \bq-tagged.
Additionally, a sideband control region, with a similar background composition to that of the signal region, is defined by adapting the hadronic \W candidate mass requirements of the signal region selection.
In the boosted (resolved) category \mj (\mjj) is required to be outside the \W boson mass window (65--105\GeV) and within the range $40 < \mj\,(\mjj) < 250\GeV$.
In the final fit to the data, the normalizations of both the \Wjets and top quark backgrounds are allowed to float freely,
with the observed yields in the control regions used to constrain the normalizations.
This background estimation procedure is applied independently in each category.

The contamination from diboson events represents 6 and 3\% of the total background in the boosted and resolved categories, respectively.
Production of \WW, \WZ, and \ZZ through \qqb annihilation is estimated directly from simulation while the \ggWW and \qqWWqq backgrounds are estimated through the reweighting of signal samples using MELA.

The DY contamination is suppressed due to the second-lepton veto.
It is estimated directly from simulation and represents between 1 and 2\% of the total background.

Contamination from nonprompt leptons in QCD multijet production is estimated from simulation to be between 1 and 2\% of the total background.
The contribution from this source is largely suppressed due to the \W candidate \pt, transverse mass, and substructure requirements.
The QCD multijet enriched samples are defined through a reversal of these requirements, allowing a test of the multijet simulation.
The resolved selection is altered by requiring $\mtw < 50\GeV$, $\mthljj < 60\GeV$, and $\ptw/\mww < 0.35$,
while for the boosted selection it is required that $\mtw < 50\GeV$, $\tauto > 0.4$, and $\ptw/\mww < 0.4$.
The QCD multijet contamination levels attained are 35 and 14\% in the boosted and resolved categories, respectively.
After subtracting the estimated prompt-lepton backgrounds, the predicted number of QCD multijet events in each category is found to agree with the data within 3\%, with the statistical uncertainties of the order of 10\%.

\begin{figure}[htbp]
\centering
\includegraphics[width=0.44\textwidth]{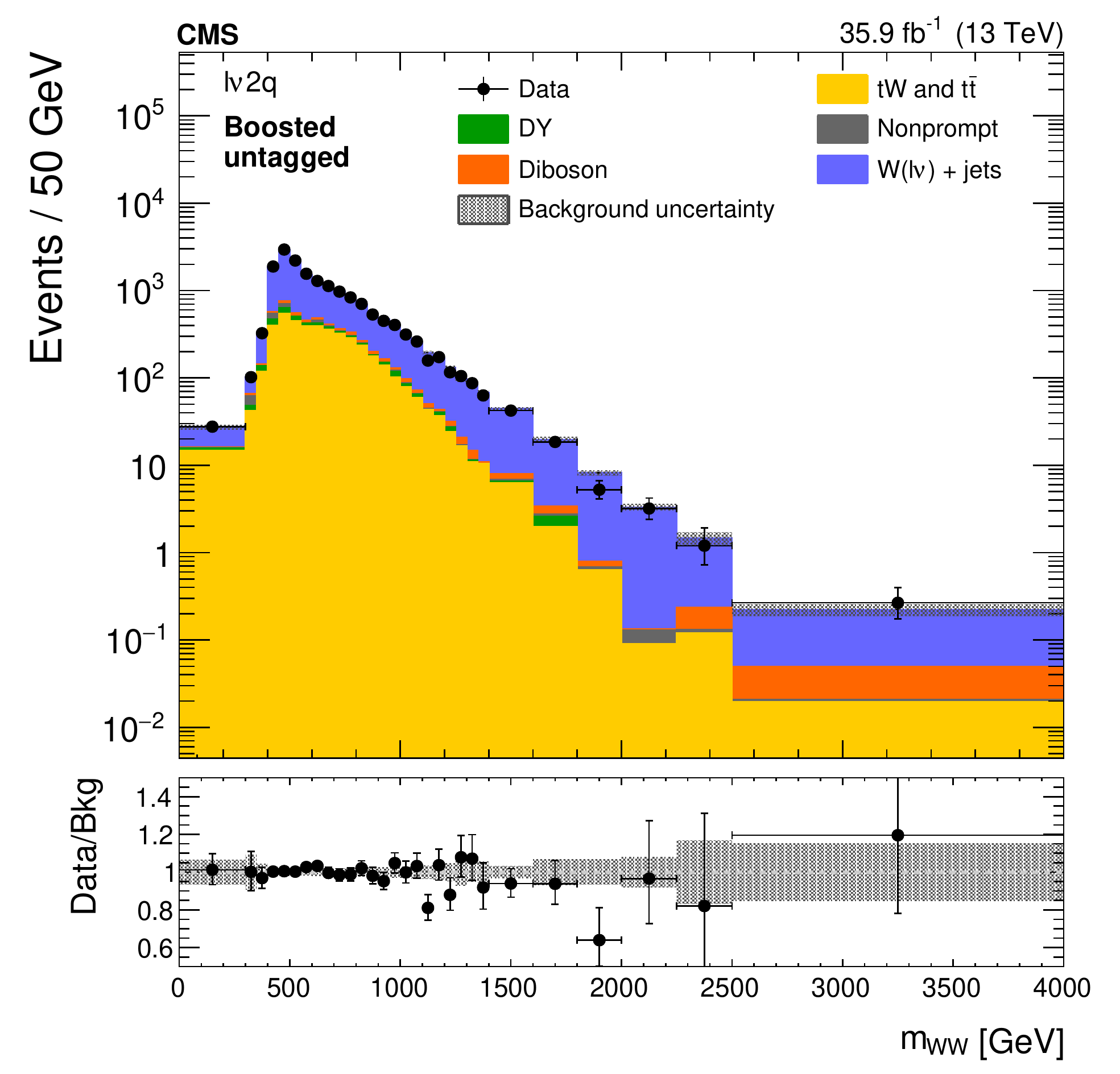}
\includegraphics[width=0.44\textwidth]{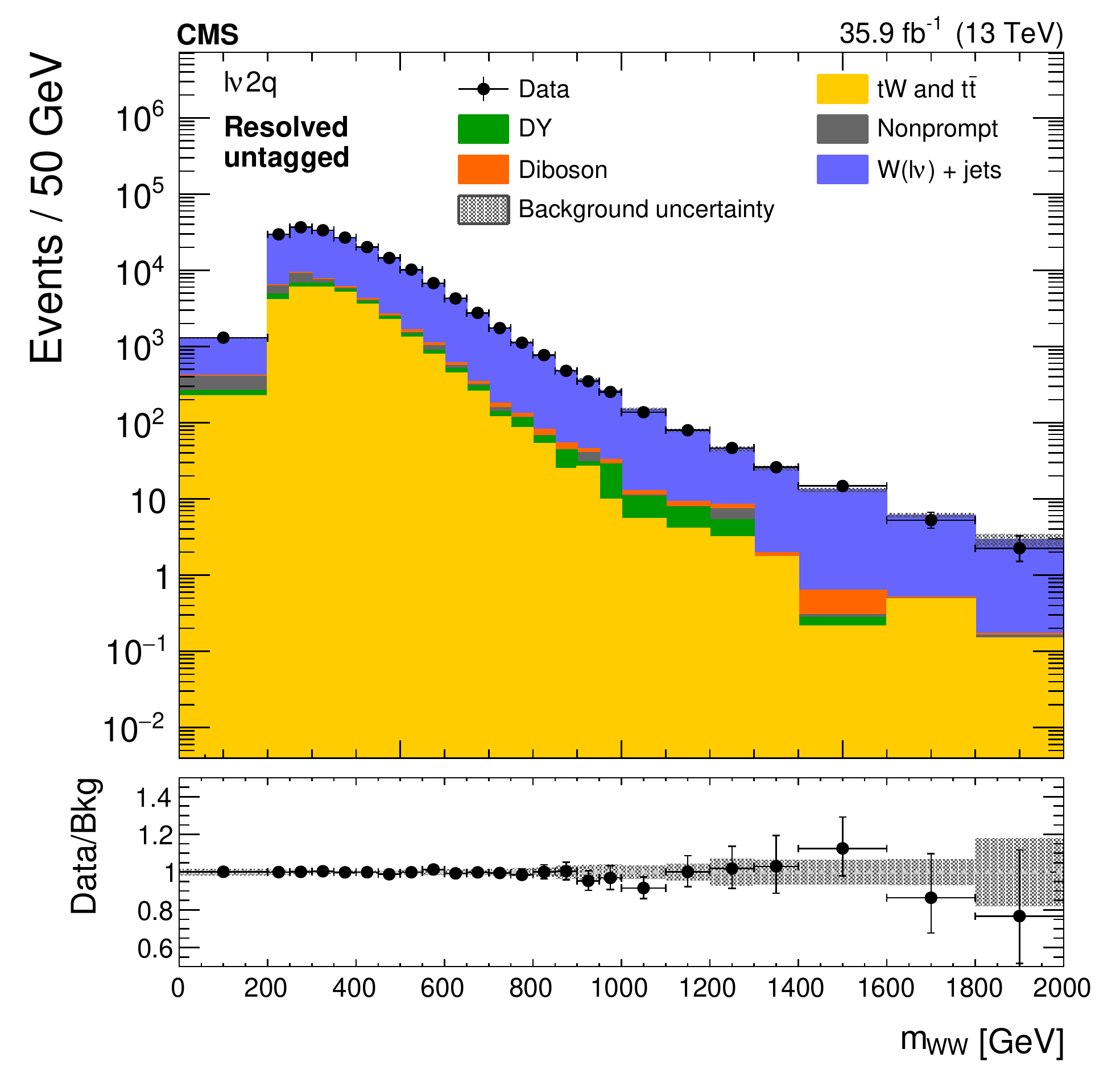} \\
\includegraphics[width=0.44\textwidth]{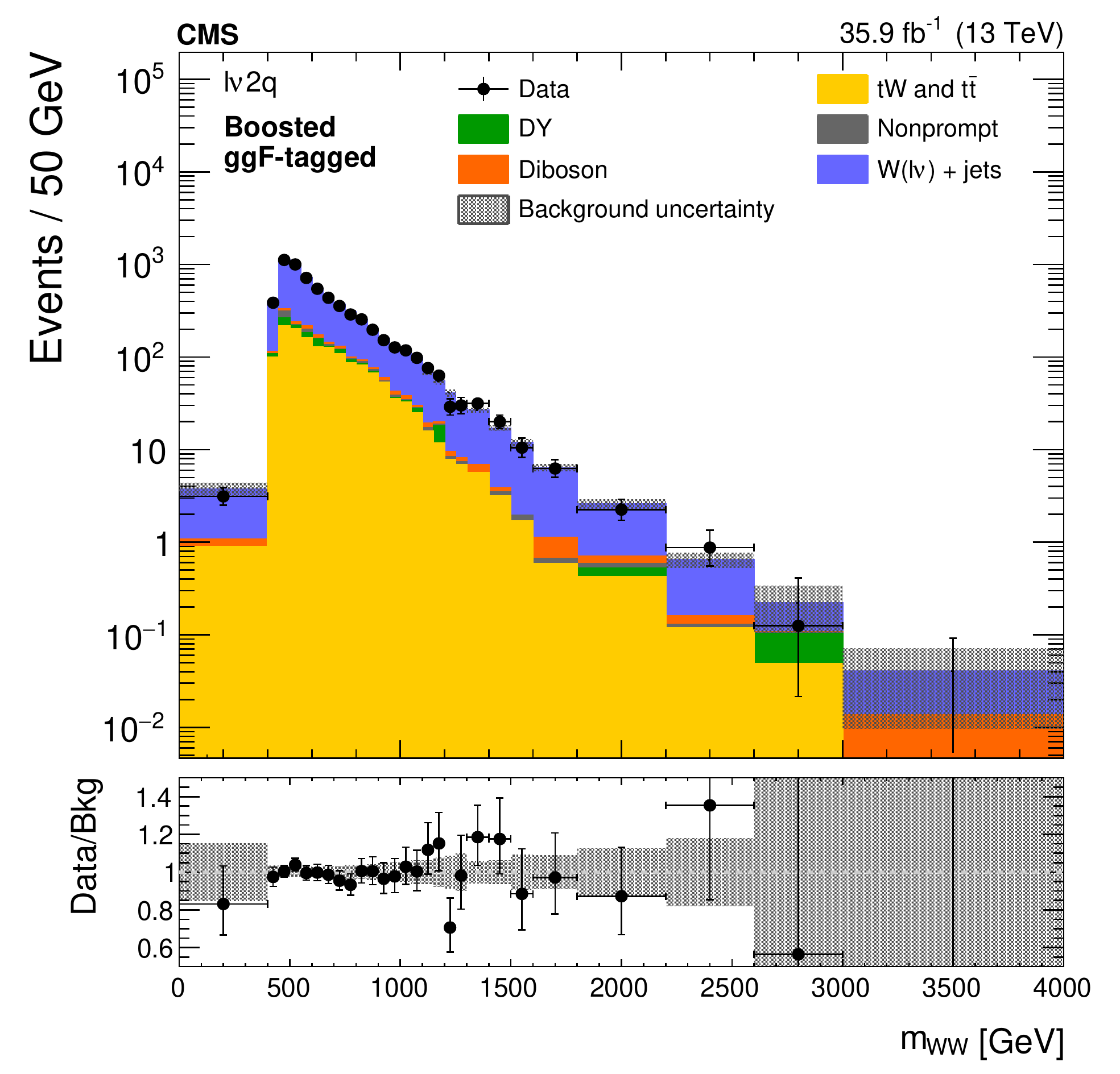}
\includegraphics[width=0.44\textwidth]{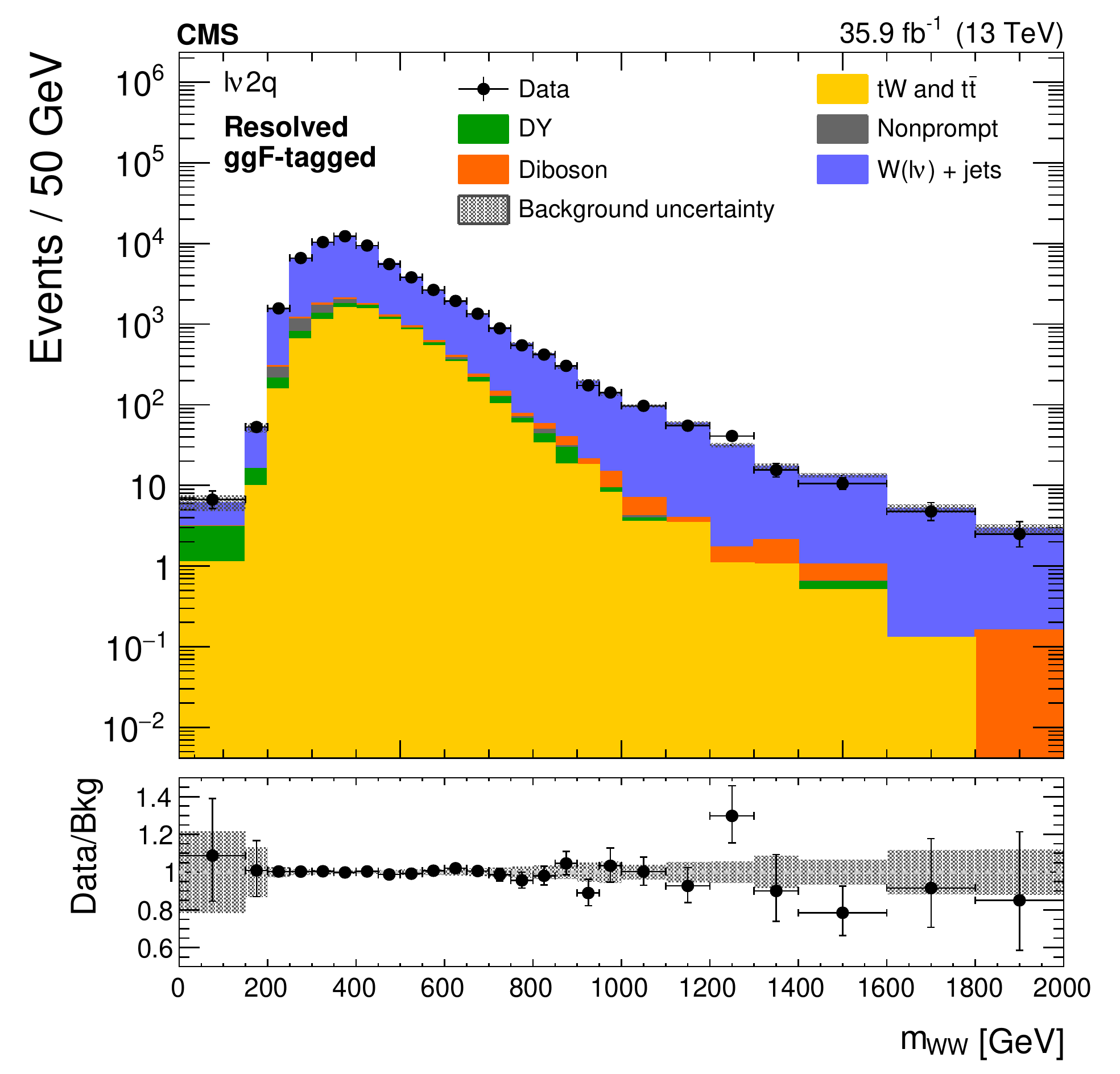} \\
\includegraphics[width=0.44\textwidth]{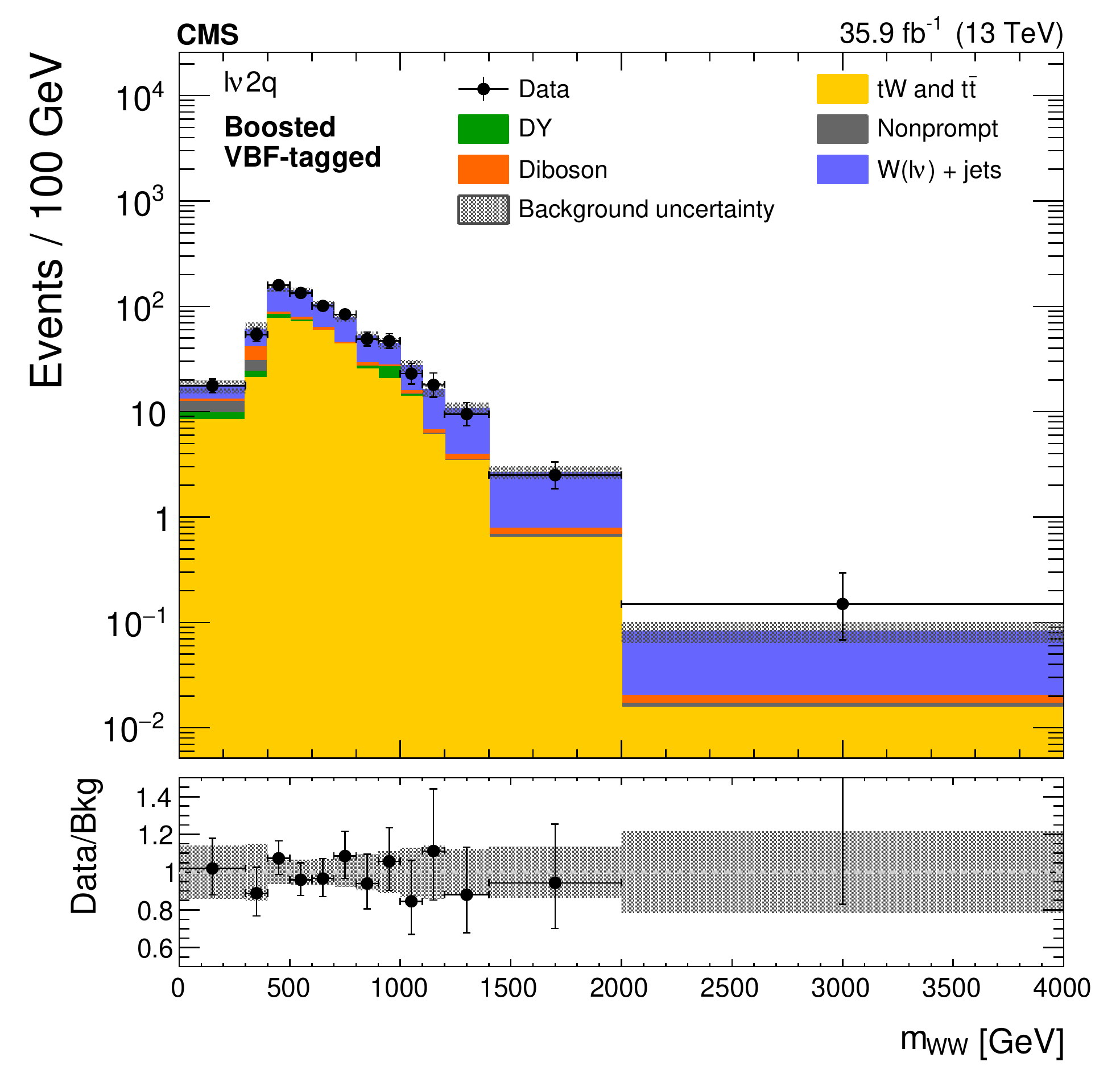}
\includegraphics[width=0.44\textwidth]{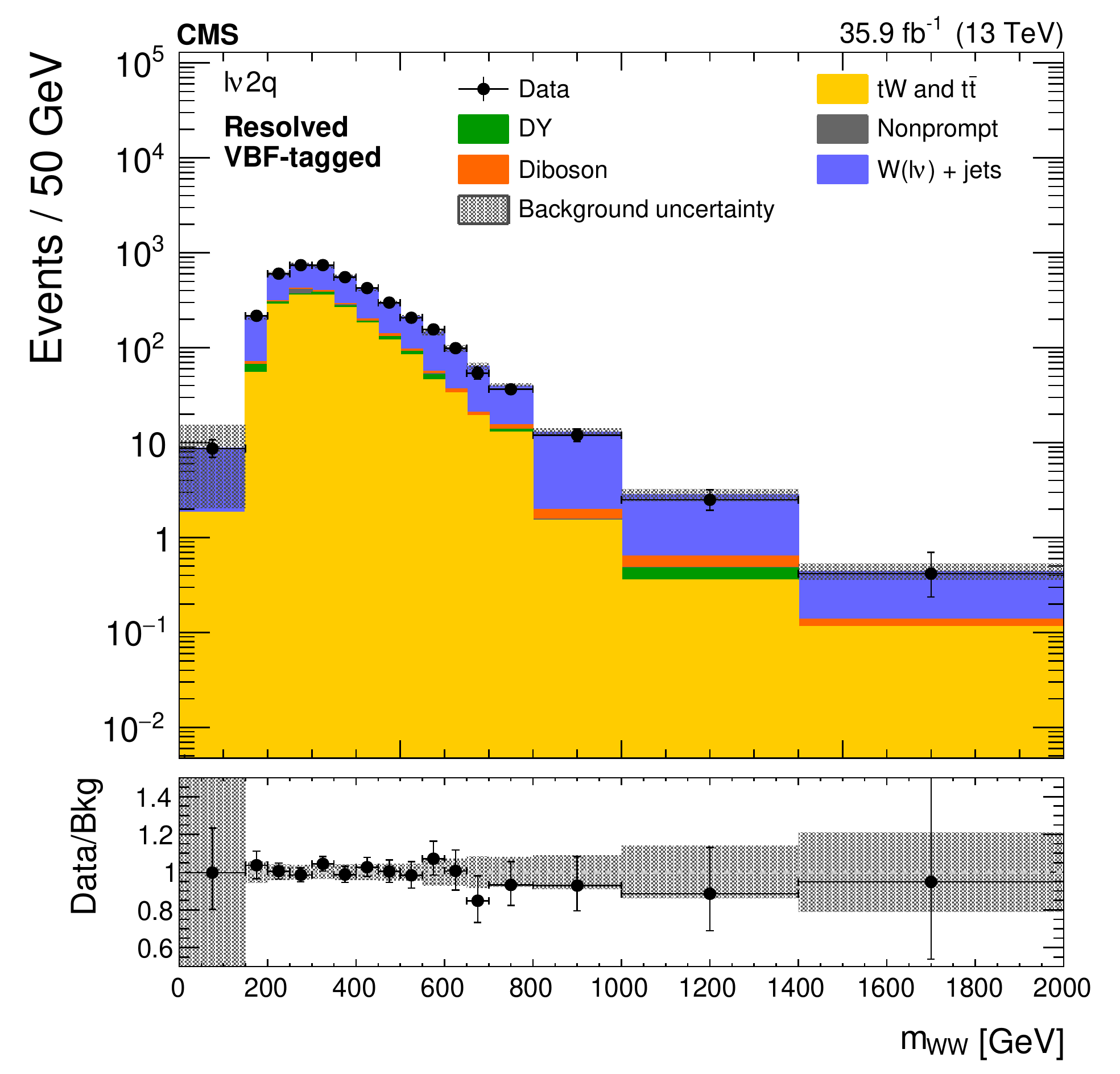} \\
\caption{The \mww distributions in the sideband control regions of the \SL boosted (left) and resolved (right) categories, after fitting the sideband data with the top quark background normalization determined using a control region.
Electron and muon channels are combined.
The points represent the data and the stacked histograms show the expected backgrounds.
The hatched area shows the combined statistical and systematic uncertainties in the background estimation.
Lower panels show the ratio of data to expected background.
Larger bin widths are used at higher \mww; the bin widths are indicated by the horizontal error bars.
}
\label{fig:HMass_SB_lnuqq}
\end{figure}

To help verify the background estimation procedure, a fit is performed to the \mww distributions in the sideband allowing the \Wjets and top quark background normalizations to float freely.
The observed yield in the top quark control region is included in the fit to help constrain the top quark background normalization.
Figure~\ref{fig:HMass_SB_lnuqq} shows the result of the fit to the sideband \mww distributions for the boosted and resolved categories.
A good level of agreement between data and the background predictions is observed.

\section{Signal extraction and systematic uncertainties}\label{sec:sys}

The methodology used to interpret the data and to combine the results from independent categories
has been developed by the ATLAS and CMS Collaborations in the context of the LHC Higgs Combination Group.
A general description of the method can be found in Refs.~\cite{Junkcls, Read1, Cowan:2010js}.

The signal extraction procedure is based on a combined binned maximum likelihood fit of the discriminant distributions with signal and background templates, performed simultaneously in all the \SL and \FL signal region categories.
Signal templates for both the \ggf and VBF production modes are included in the fit, with a number of hypotheses for \fvbf considered.
The various control regions used to constrain the dominant backgrounds are included in the form of single bins, representing the number of events in each control region.
The dominant background normalizations are initially unconstrained and are determined during the fit.
After fitting the data the uncertainties on the \WW, top quark and DY background normalizations in the \FL categories are in the range 6--45\% , 3--5\%, and 5--20\%, respectively.
In the \SL categories, the corresponding uncertainties on the \Wjets and top quark background normalizations are in the range 7--10\% and 4--20\%, respectively.
The remaining systematic uncertainties are represented by individual nuisance parameters with a log-normal model used for normalization uncertainties and a Gaussian model used for shape uncertainties.
For each source of uncertainty, the correlations between different categories, and different signal and background processes, are taken into account.
Uncertainties arising from limited number of events in the MC simulated samples are included for each bin
of the discriminant distributions, in each category independently, following the Barlow--Beeston approach~\cite{Barlow:1993dm}. 
Depending on the category, the statistical uncertainties due to the MC simulated sample sizes on the background and signal normalisations are in the range 1--8\%.

The theoretical sources of uncertainty considered include the effect of PDFs and the strong coupling constant \alpS, and the effect of missing higher-order corrections via variations of the renormalization and factorization scales. Acceptance uncertainties are evaluated for signal and background by varying the PDFs and \alpS within their uncertainties~\cite{Butterworth:2015oua}, and by varying the factorization and renormalization scales up and down by a factor of two~\cite{Cacciari:2003fi}. Depending on the process and the category, the PDF uncertainties in the signal and background yields amount to 1--7\%, while those of the renormalization and factorization scales are within 1--18\%. The PDF, and the renormalization and factorization scales uncertainties in the signal cross section, computed by the LHC Higgs Cross Section Working Group~\cite{deFlorian:2016spz}, are also considered and amount to 2--16\% and 0.2--9\%, respectively, depending on the resonance mass and production mechanism.

Effects due to experimental uncertainties are studied by applying a scaling and/or smearing of certain variables of the physics objects in the simulation,
followed by a subsequent recalculation of all the correlated variables.
The uncertainty in the measured luminosity is 2.5\% for data collected during 2016~\cite{CMS:2017sdi}.
The trigger efficiency uncertainties are approximately 1 and 2\% for the \SL and \FL final states, respectively.
Lepton reconstruction and identification efficiency uncertainties vary between 1 and 3\%, while the muon momentum and electron energy scale uncertainties amount to 0.1--1.0\% each.
Depending on the process and the category, the jet energy scale uncertainties are in the range 1--10\%.
The \ptmiss uncertainty is taken into account by propagating the corresponding uncertainties in the leptons and jets and amounts to 0.1--1\%.
The scale factors correcting the \bq tagging efficiency and mistag rate are varied within their uncertainties with resulting uncertainties of 0.1--5\% depending on the process and the category.
This systematic uncertainty affects the top quark control regions and the signal regions in an anticorrelated way.

In addition, for each final state there are channel-specific uncertainties which are now discussed.

\subsection{ \texorpdfstring{\PX \ra \FL}{2l2nu} }

A conservative 30\% uncertainty in the normalization of the instrumental background arising from nonprompt leptons in \Wjets production is estimated by varying the jet \pt threshold in the dijet control sample used in the background prediction procedure, and from propagation of the statistical uncertainties in the measured lepton misidentification probabilities.
Uncertainties of 3--10\% due to the \ptww reweighting are evaluated by varying the factorization and renormalization scales up and down by a factor of two, and by varying the resummation scale.
The UE uncertainty for the \WW background is estimated by comparing two different UE tunes,
while the PS modeling uncertainty is estimated by comparing samples interfaced
with different PS models, as described in Section~\ref{sec:mcdata}.
The combined effect is evaluated to be 5-10\%.
A dedicated nuisance parameter for the linear \ptmiss correction in the same-flavour DY control region is introduced.
The uncertainty is 0.2--1\%, estimated with the maximum and minimum best fit lines of the linear fit used to derive the correction.
The categorization of events based on jet multiplicity introduces additional signal uncertainties related to higher-order corrections.
These uncertainties are associated with the \ggf production mode and are evaluated independently following the method described in Ref.~\cite{Boughezal:2013oha} and are about 5\% for the 0-jet, 10\% for the 1-jet, and 20\% for the 2-jet and VBF categories.

\subsection{ \texorpdfstring{\PX \ra \SL}{lnu2q} }

The diboson and DY production cross sections are each assigned an uncertainty of 10\% based on the level of agreement between theoretical predictions and cross section measurements at CMS using 13\TeV data~\cite{Sirunyan:2019bez, Sirunyan:2018cpw}.
An uncertainty of 10\% in the normalization of the background arising from nonprompt leptons in QCD multijet production is assigned based on the observed level of agreement between data and simulation in QCD multijet enriched samples.
The impact of the jet energy resolution uncertainty is about 0.3--2\%, depending on the process and the category.
For \W-tagged jets the \mj scale and resolution uncertainties are evaluated to be 0.1--1 and 2--5\%, respectively.
The \tauto scale factor correcting the boosted \W tagging efficiency has an associated uncertainty of 6\%. Since this is measured in \ttbar events using jets with a typical \pt of 200\GeV, an uncertainty of 1--13\% in the extrapolation to the higher-\pt regime of the high-mass signal is also included.

A summary of the systematic uncertainties included for the \SL and \FL final states are shown in Table~\ref{table-syst}.

\begin{table}[htbp]
\centering
\topcaption{
Summary of systematic uncertainties, quoted in percent, affecting the normalization of the background and signal samples.
The uncertainties on the \WW, top quark and DY (\Wjets and top quark) background estimates in the \FL (\SL) categories have been determined during the fit to the data.
The numbers shown as ranges represent the uncertainties for different processes and categories.
Missing values represent uncertainties either estimated to be negligible ($<$0.1\%), or not applicable in a specific channel.
Those systematic uncertainties found to affect the shape of kinematic distributions are labeled with *.\label{table-syst}
}
\cmsTable{
\begin{tabular}{>{\hspace{3mm}}lccc}
\hline
 \multicolumn{1}{l}{Source of uncertainty}  & \PX \ra \WW \ra \FL              & \PX \ra \WW \ra \SL   & \PX \ra \WW \ra \SL   \\
                        &                                 & Resolved             & Boosted   \\
\hline
 \multicolumn{1}{l}{Experimental sources} & \multicolumn{3}{c}{}    \\ [\cmsTabSkip]
Integrated luminosity            &   2.5\%    &   2.5\%     &  2.5\% \\
Lepton trigger*                  &   2\%      &   1\%       &  1\% \\
Lepton reconstruction $\&$ ident.*   &  1--3\%     &   1--2\%     &  1--2\% \\
Electron energy scale*           &  0.1--1\%   &   0.2--1\%   & 0.1--1\% \\
Muon momentum scale*             &  0.1--1\%   &   0.1--1\%   & 0.1--1\% \\
Jet energy scale*                &  1--10\%    &   1--6\%     & 1--3\% \\
Jet energy resolution*           &   \NA      &   0.5--2\%   & 0.3--1\% \\
\ptmiss*                            &  0.1--1\%   &   1--3\%     & 0.1--1\% \\
\bq tagging/mistag*              &  0.1--5\%   &   0.1--1\%   & 0.1--1\% \\
\W tagging (\tauto)              &  \NA  &  \NA      &  6\%   \\
\W tagging (extrapolation)       &  \NA  &  \NA      &  1--13\%   \\
\W \mj scale          &  \NA  &  \NA      &  0.1--1\%   \\
\W \mj resolution     &  \NA  &  \NA      &  2--5\% \\ [\cmsTabSkip]
 \multicolumn{1}{l}{Background estimates} & \multicolumn{3}{c}{}   \\ [\cmsTabSkip]
\WW           &  6--45\%  &  10\%         &   10\%    \\
top quark     &  3--5\%   &  7--9\%        &   8--10\%   \\
\Wjets        &  30\%    &  5--11\%       &   4--20\%   \\
QCD multijet  &  \NA     &  10\%         &   10\%   \\
DY            &  5--20\%  &  10\%         &   10\%    \\ [\cmsTabSkip]
 \multicolumn{1}{l}{Theoretical sources}  & \multicolumn{3}{c}{}    \\ [\cmsTabSkip]
PDF and \alpS (acceptance)*                       & 1--4\%   &  1--4\%   &  1--7\% \\
Renorm./factor. scales (acceptance)*    & 1--6\%   &  1--18\%  &  1--18\% \\
PDF and \alpS ($\sigma_{\PX}$)                      & 2--16\%  &  2--4\%  &  2--16\% \\
Renorm./factor. scales ($\sigma_{\PX}$)   & 0.2--9\% &  0.2--4\% & 0.2--9\% \\
Jet multiplicity categorization ($\sigma_{\Pg\Pg \ra \PX}$)* & 5--20\% &    \NA &  \NA     \\
\WW \ptww reweighting*      & 3--10\% &  \NA     &  \NA            \\
\WW UE $\&$ PS                & 5--10\% &  \NA    &  \NA              \\
DY \ptmiss reweighting*        & 0.2--1\% &  \NA    &  \NA   \\ [\cmsTabSkip]
 \multicolumn{1}{l}{Other sources}  & \multicolumn{3}{c}{}    \\ [\cmsTabSkip] 
MC statistics*                       & 1--5\%   &  1--8\%   &  1--5\% \\
\end{tabular}
}
\end{table}

\section{Results}\label{sec:results}

\begin{figure}[htbp]
\centering
\includegraphics[width=0.495\textwidth]{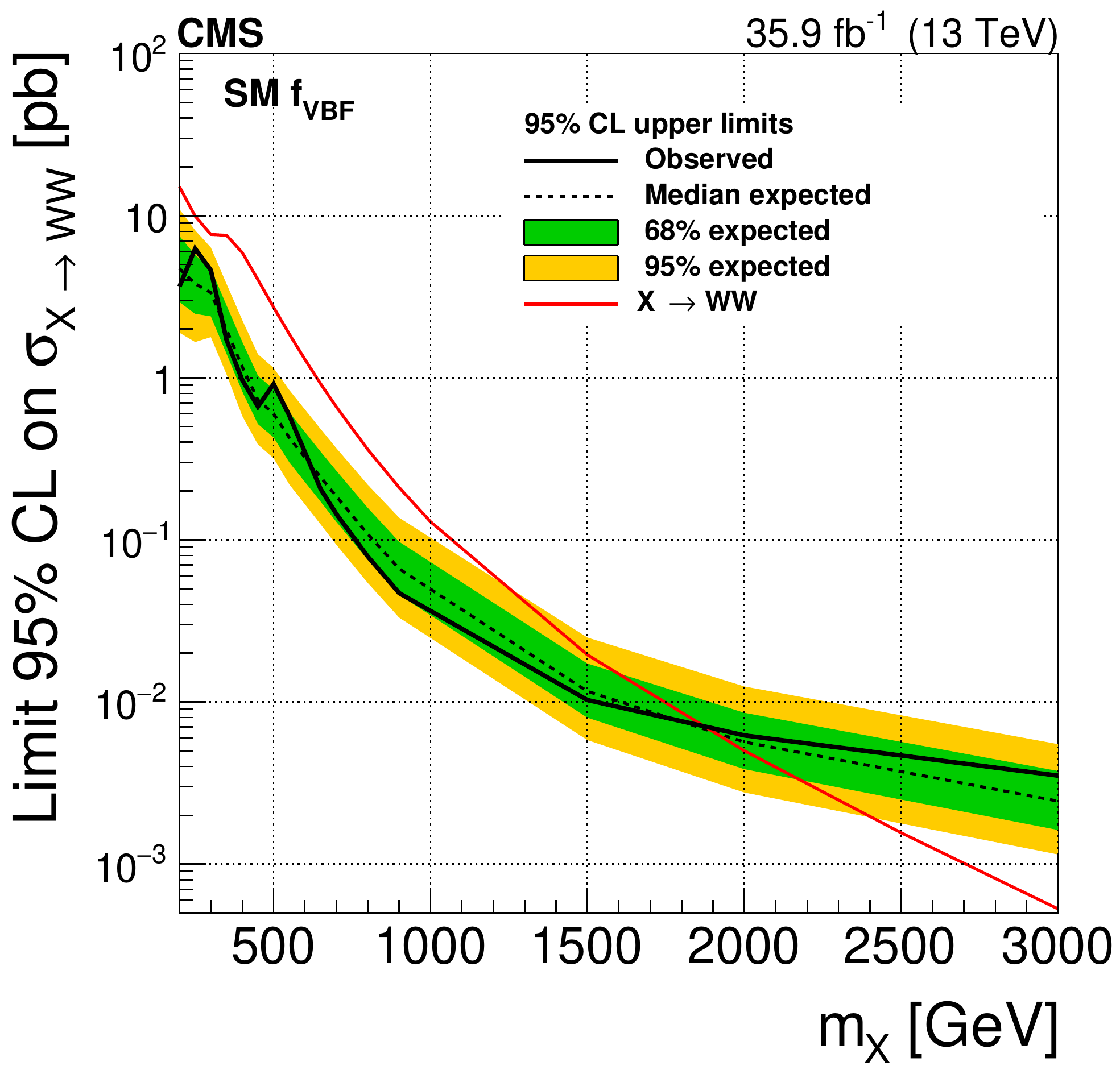}
\includegraphics[width=0.495\textwidth]{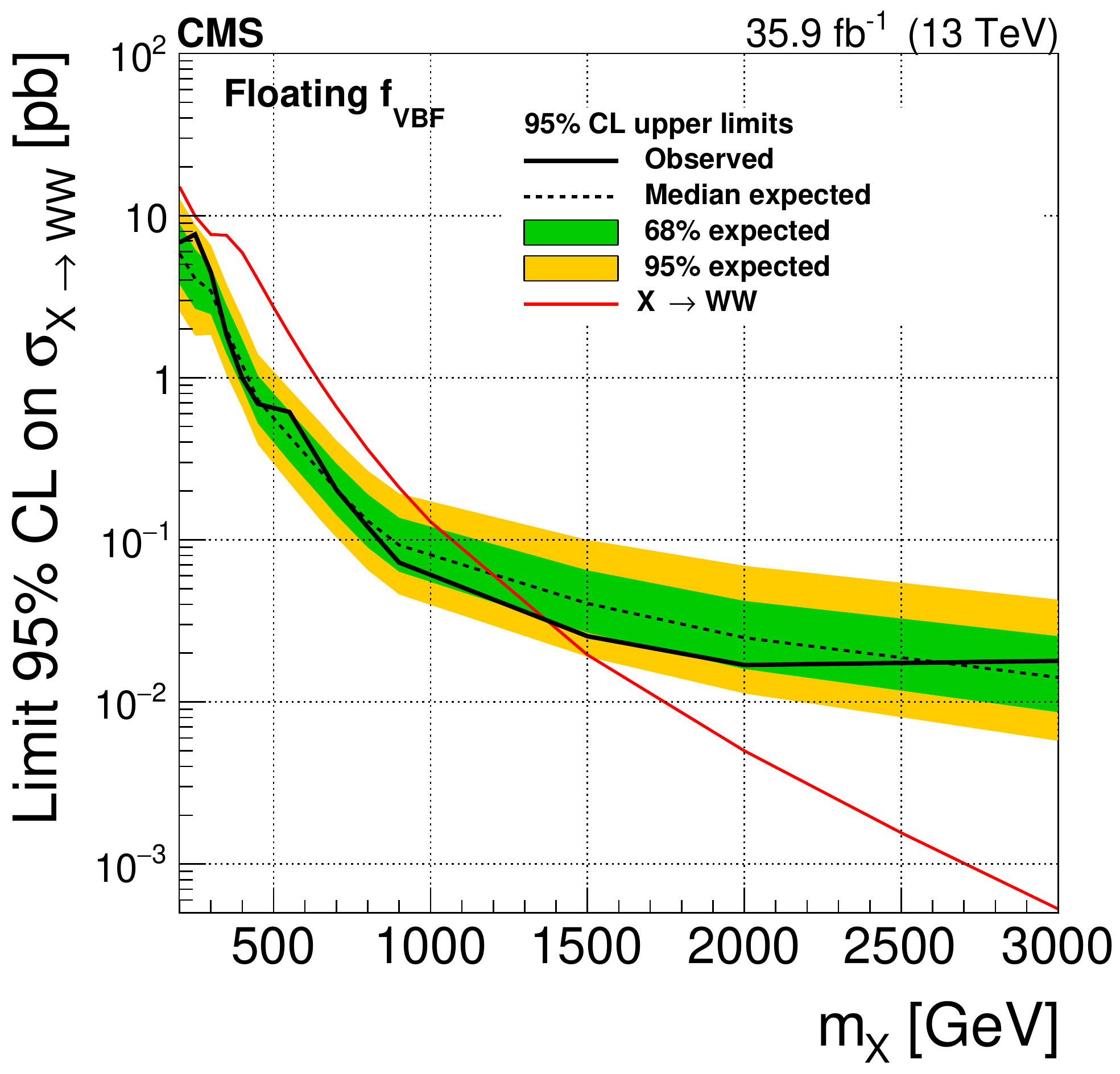}\\
\includegraphics[width=0.495\textwidth]{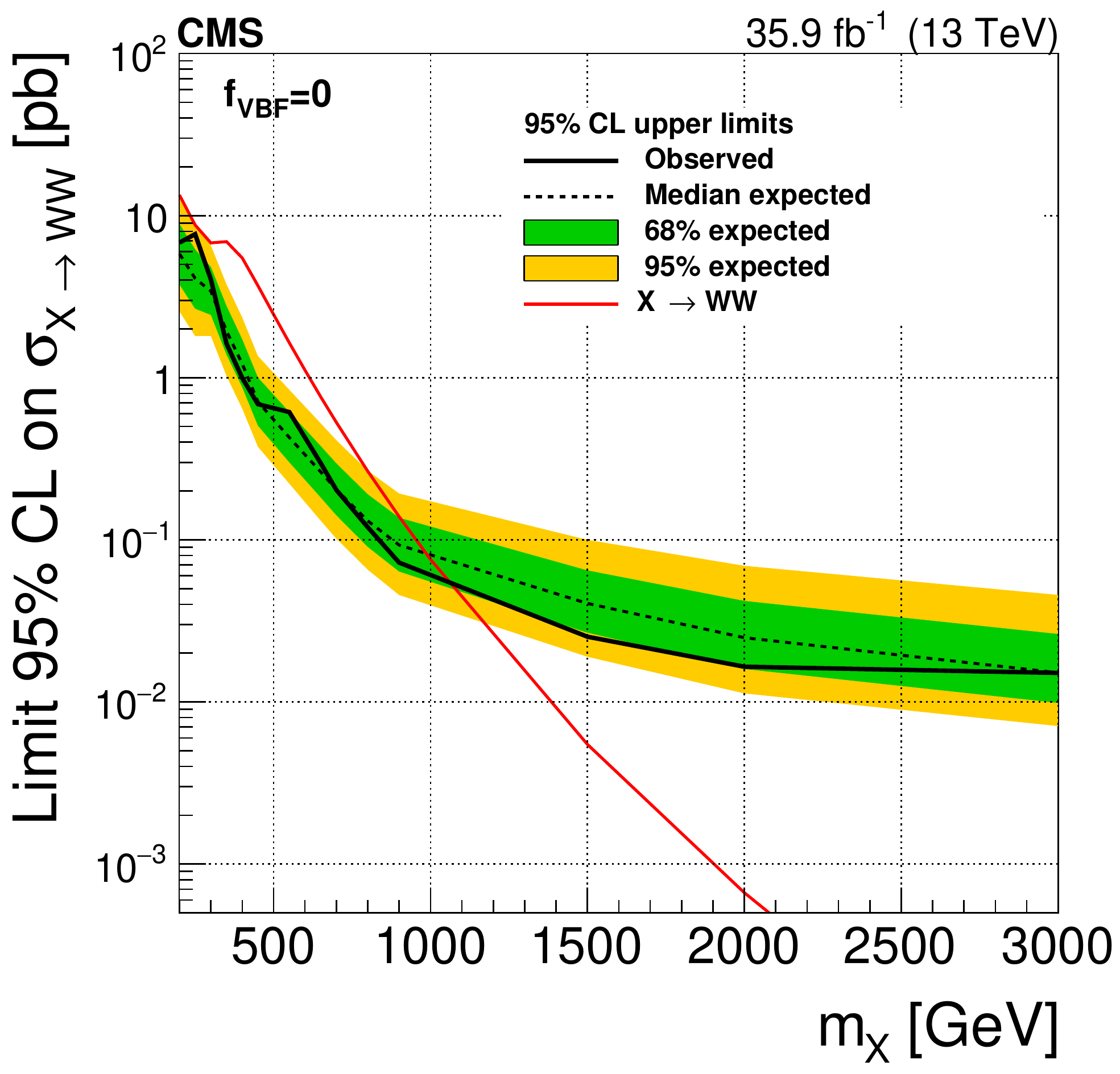}
\includegraphics[width=0.495\textwidth]{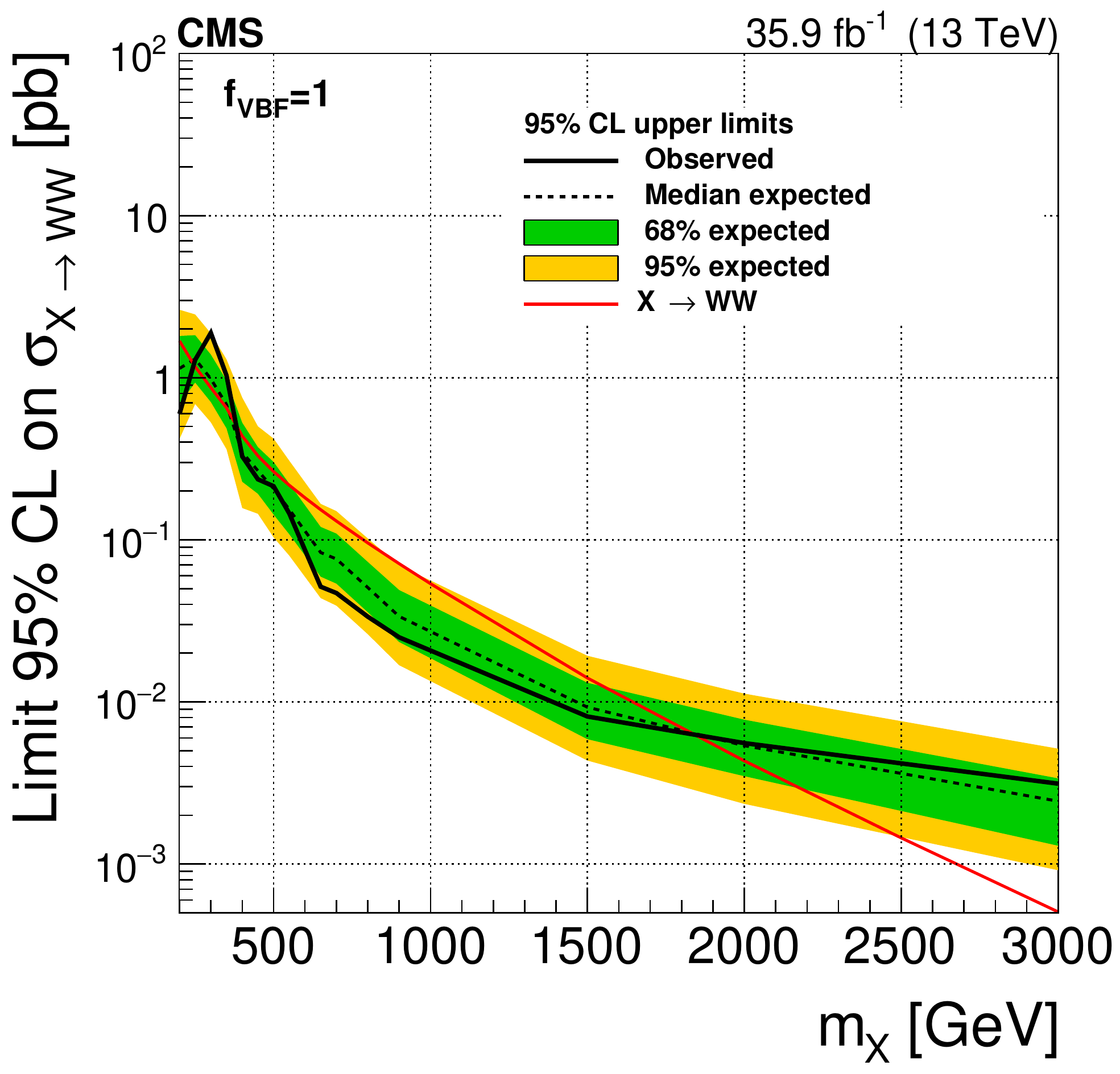}

\caption{ Expected and observed exclusion limits at 95\% \CL on the \PX cross section times branching fraction to \WW for a number of \fvbf hypotheses.
          For the SM \fvbf (upper left) and floating \fvbf (upper right) cases the red line represents the sum of the SM cross sections for \ggf and VBF production,
          while for the $\fvbf = 0$ (lower left) and the $\fvbf = 1$ (lower right) cases it represents the \ggf and VBF production cross sections, respectively.
          The black dotted line corresponds to the central expected value while the yellow and green bands represent the 68 and 95\% \CL uncertainties, respectively.
        }
    \label{fig:limit_xsec}
\end{figure}

No evidence for an excess of events with respect to the SM predictions is observed.
Upper exclusion limits at 95\% confidence level (\CL) on the \PX cross section times branching fraction of the decay to two \W bosons are evaluated for masses between 0.2 and 3.0\TeV using the asymptotic modified frequentist method (\CLs)~\cite{Junkcls, Read1, Cowan:2010js}. A number of hypotheses for \fvbf have been investigated by setting this fraction to the SM value, by allowing it to float, and by setting $\fvbf = 0$ and 1. The expected and observed exclusion limits for the full combination of the \FL and \SL analyses are shown in Fig.~\ref{fig:limit_xsec}. For signals below ${\approx}800\GeV$, the sensitivity of the \FL final state is dominated by the different-flavour channel, while at higher masses the same- and different-flavour channels have similar sensitivities.
For the \SL final state, the sensitivity is dominated by the boosted channel for signals above ${\approx}400\GeV$, while at lower masses the resolved channel dominates. 
Comparing the two final states, the \FL sensitivity is dominant up to ${\approx}400\GeV$, while at higher masses the \SL final state is more sensitive by a factor of approximately two. Comparing the excluded cross section values to the expectations from theoretical calculations, a \PX signal is excluded up to 1870 (1370)\GeV with \fvbf set to the SM value (\fvbf allowed to float).
A \PX signal is excluded up to 1060\GeV for the $\fvbf = 0$ hypothesis, while the mass ranges 200--245 and 380--1840\GeV are excluded for the $\fvbf = 1$ hypothesis.

\begin{figure}[htbp]
\centering
\includegraphics[width=0.495\textwidth]{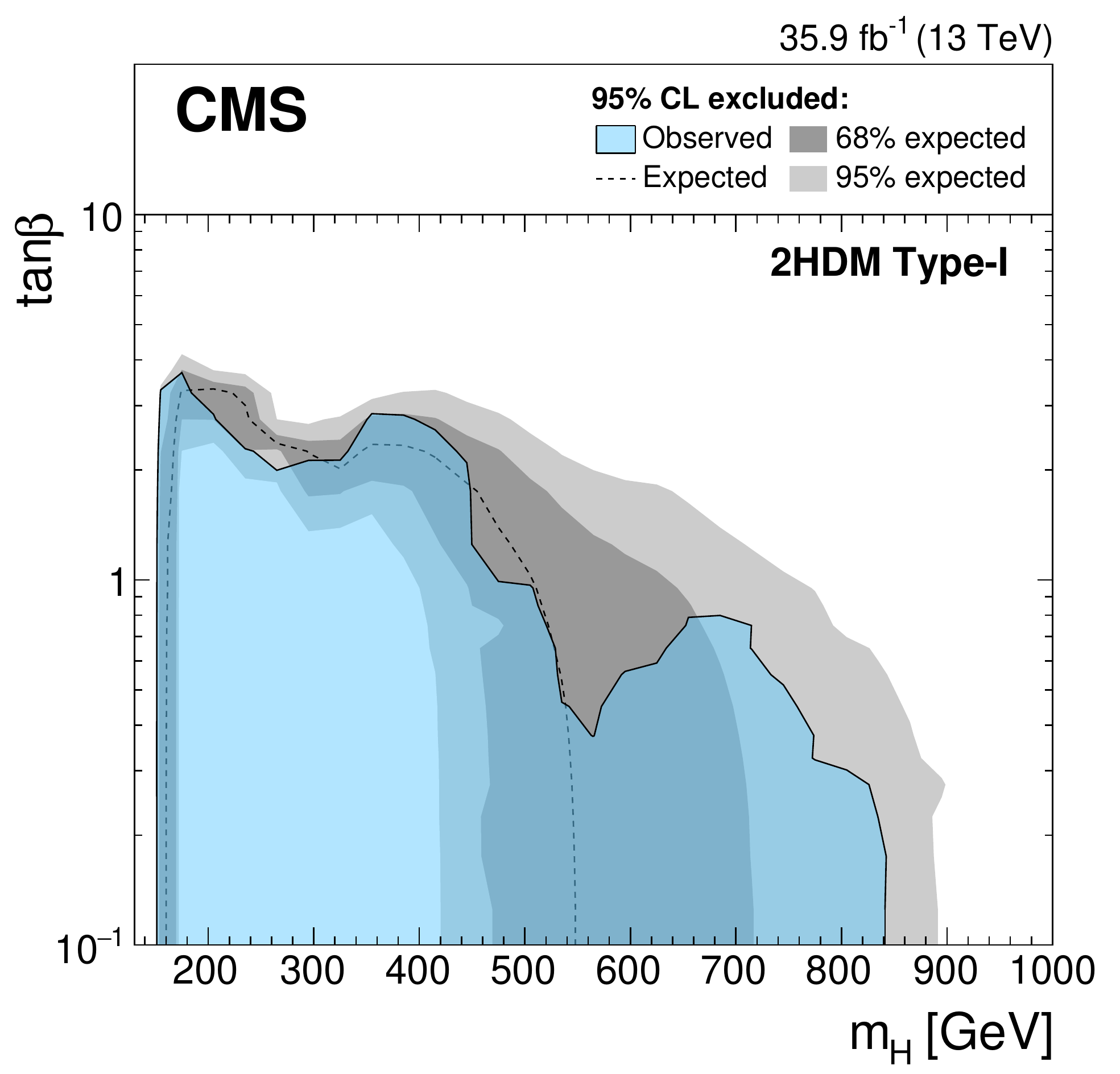}
\includegraphics[width=0.495\textwidth]{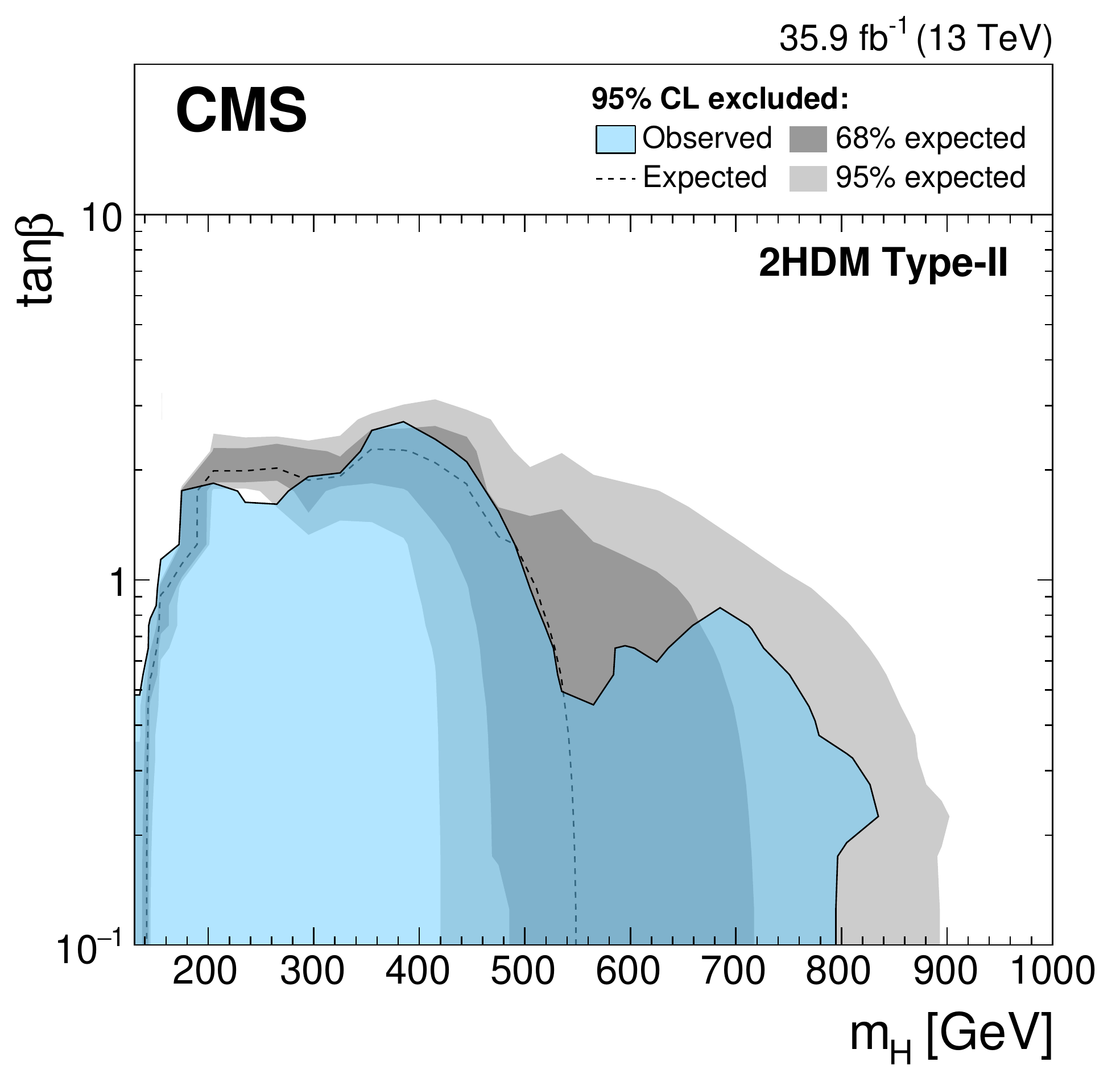}

\caption{Expected and observed 95\% \CL upper limits on \tanbeta as a function of \mH for a Type-\RNum{1} (left) and Type-\RNum{2} (right) 2HDMs. It is assumed that $\mH = \ma = \mHpm$ and $\cosba = 0.1$.
The expected limit is shown as a dashed black line while the dark and light gray bands indicate the 68 and 95\% \CL uncertainties, respectively. The observed exclusion contour is indicated by the blue area.
}
    \label{fig:2HDM}
\end{figure}

\begin{figure}[htbp]
\centering
\includegraphics[width=0.495\textwidth]{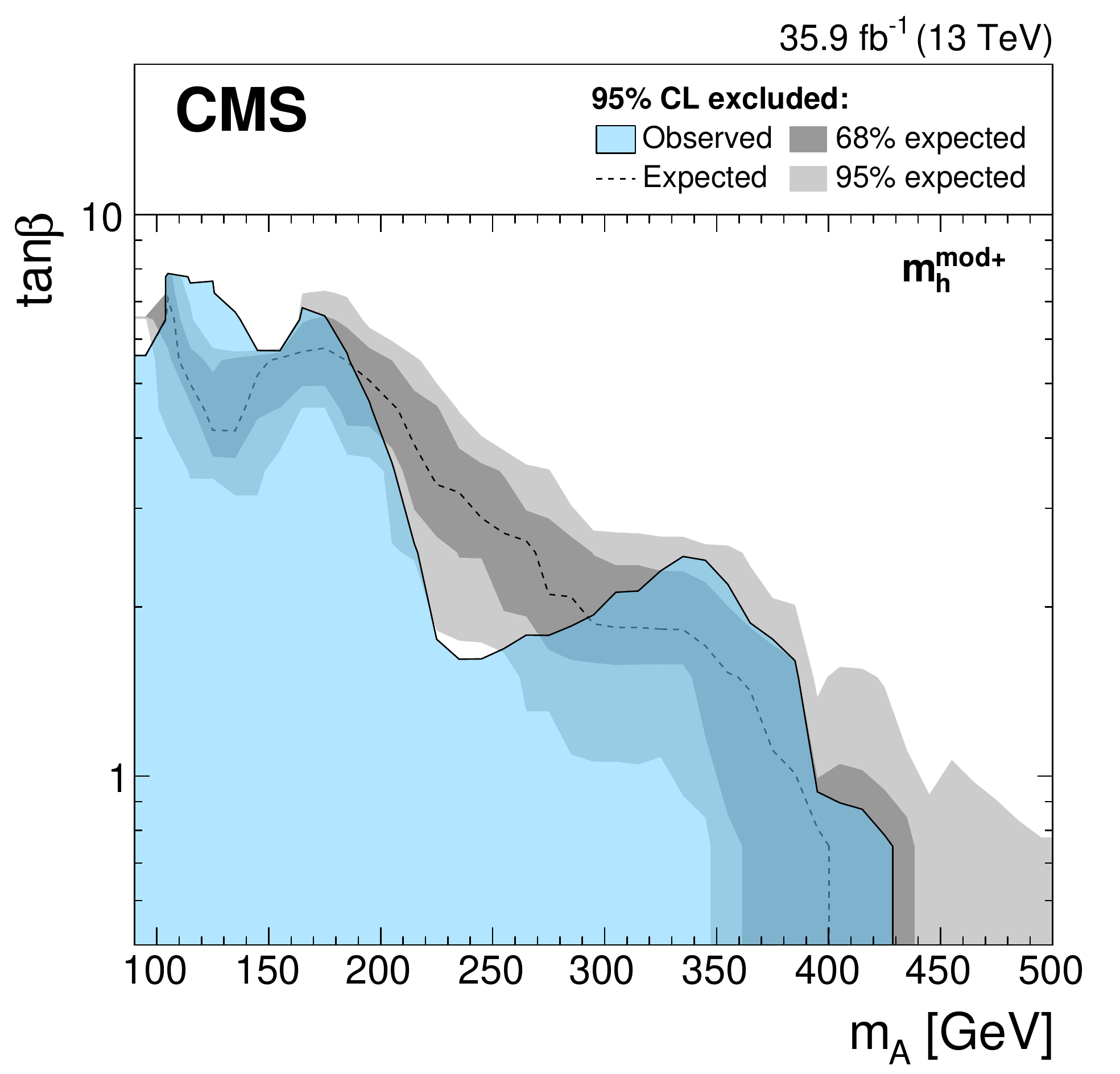}
\includegraphics[width=0.495\textwidth]{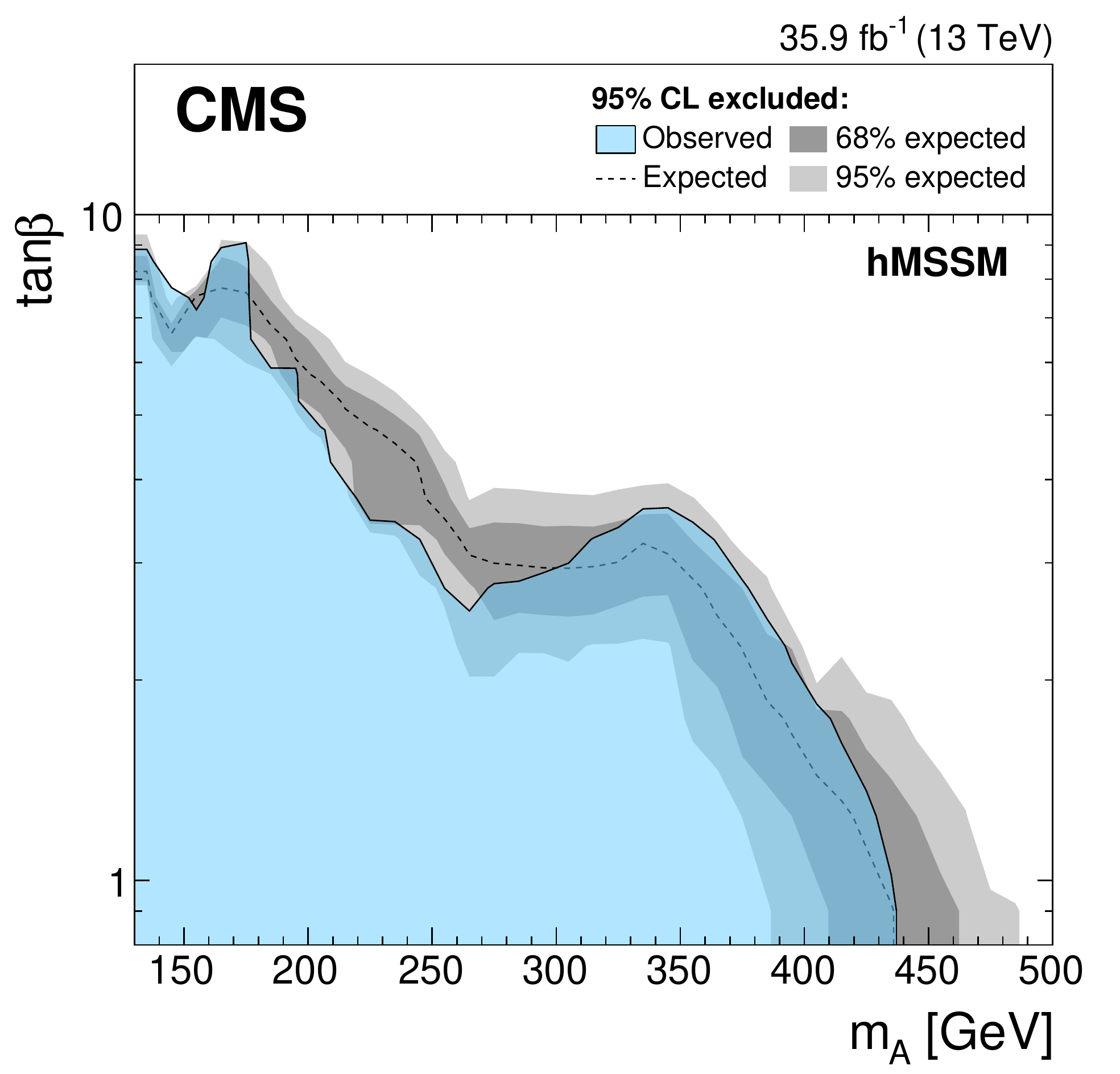}

\caption{
Expected and observed 95\% \CL upper limits on \tanbeta as a function of \ma for the \mmodp (left) and hMSSM (right) scenarios.
The expected limit is shown as a dashed black line while the dark
and light gray bands indicate the 68 and 95\% \CL uncertainties, respectively.
The observed exclusion contour is indicated by the blue area.
}
    \label{fig:MSSM}
\end{figure}

\begin{figure}[htbp]
\centering
\includegraphics[width=0.495\textwidth]{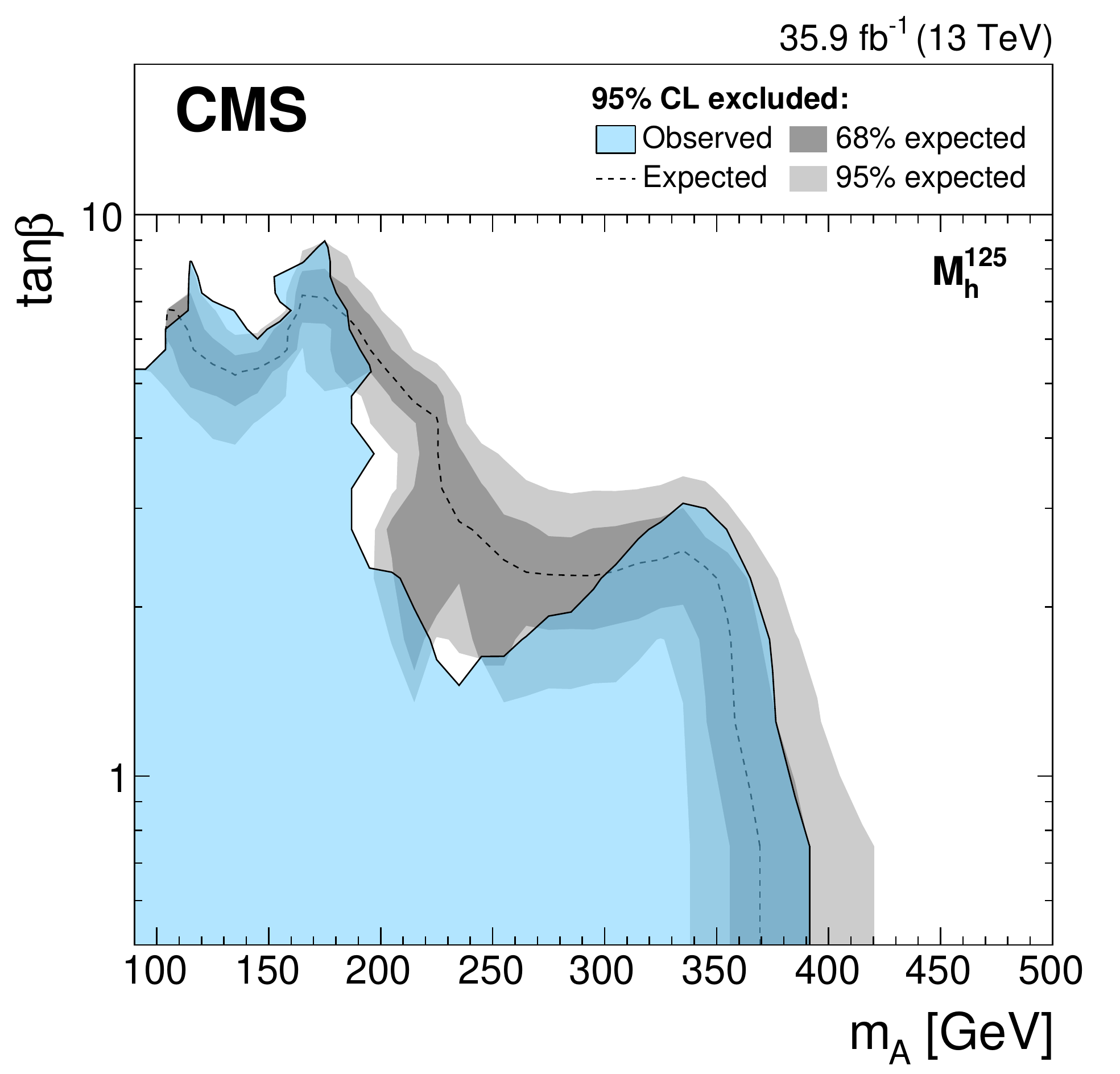}
\includegraphics[width=0.495\textwidth]{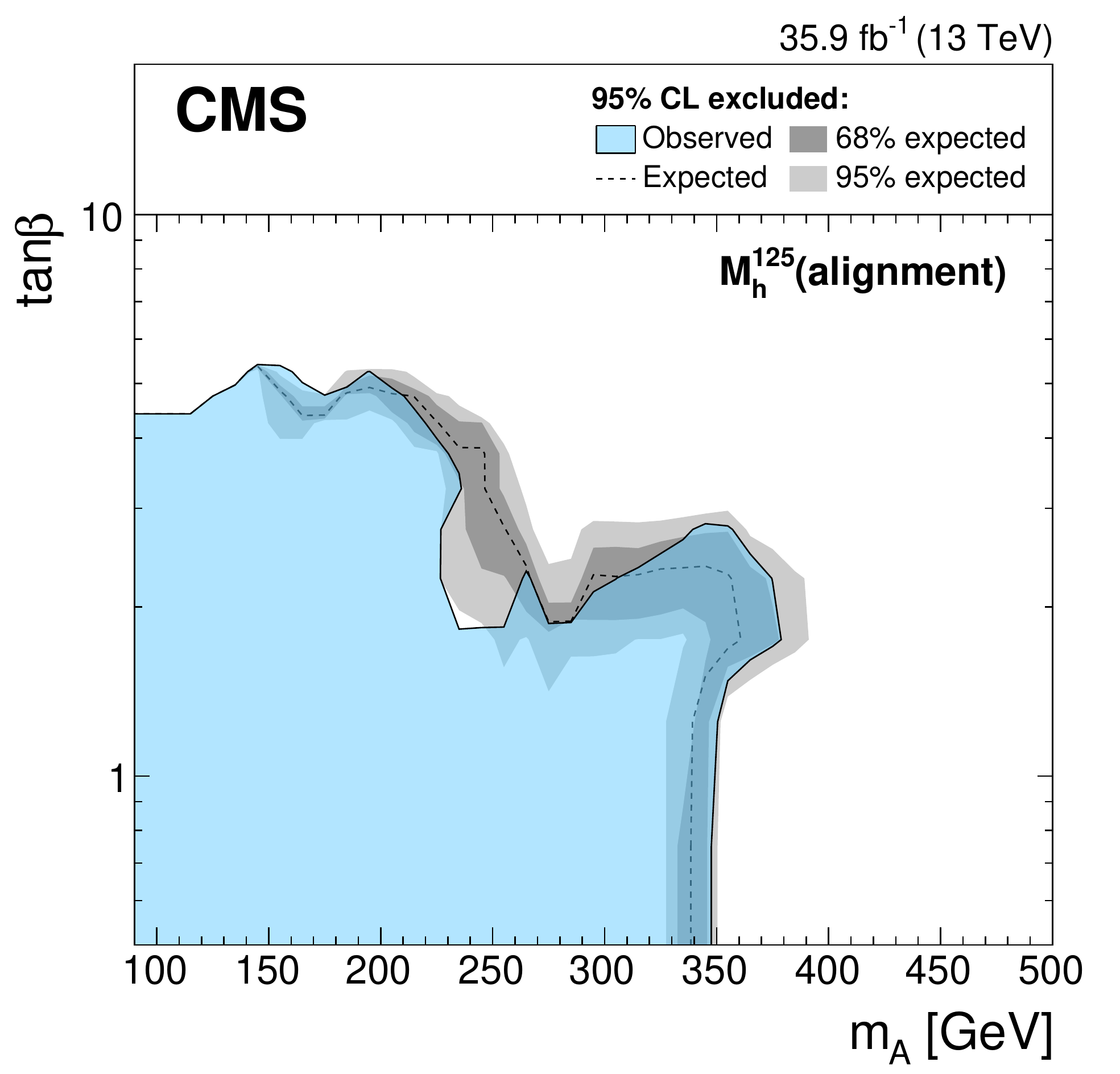} \\
\includegraphics[width=0.495\textwidth]{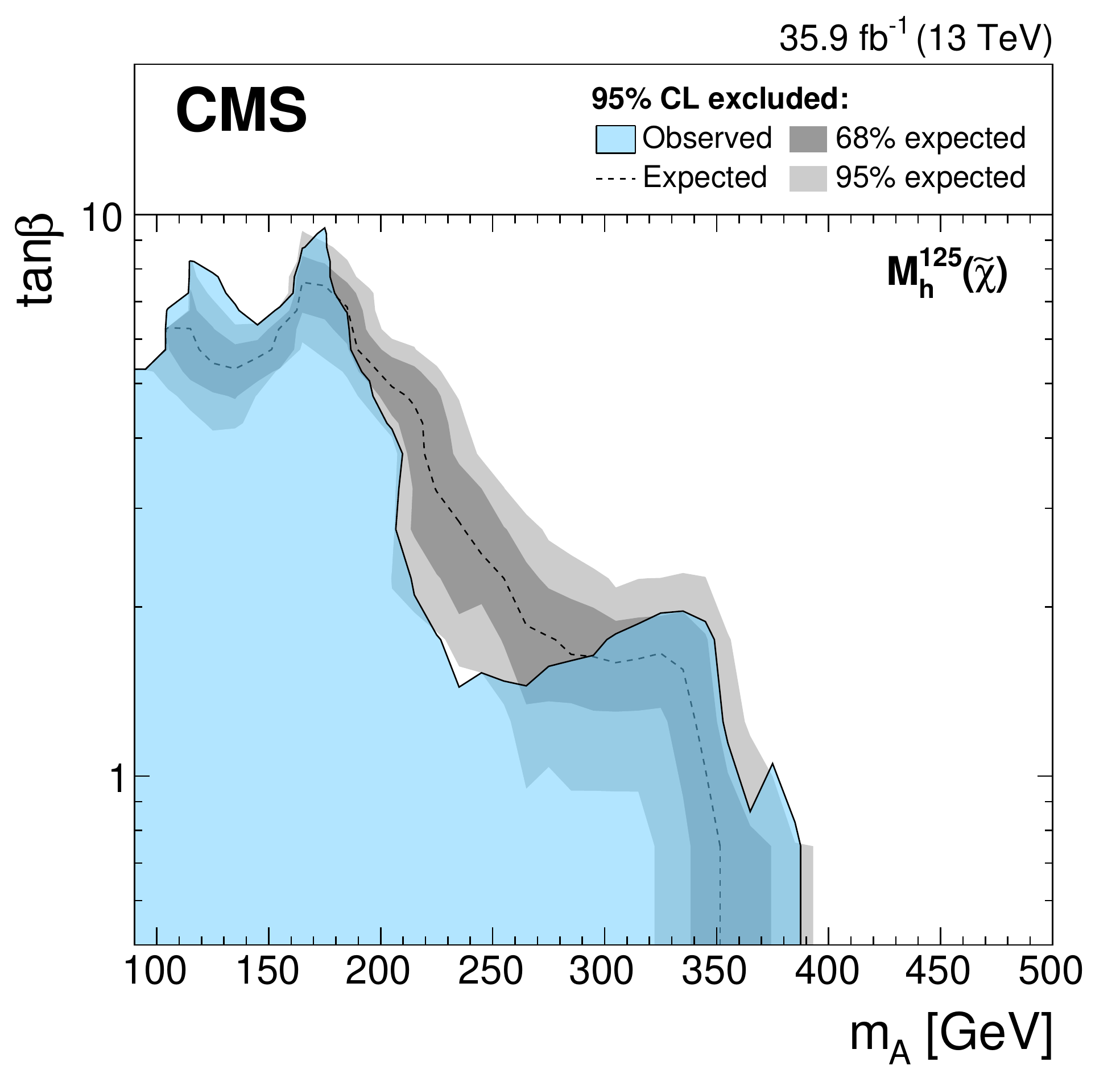}
\includegraphics[width=0.495\textwidth]{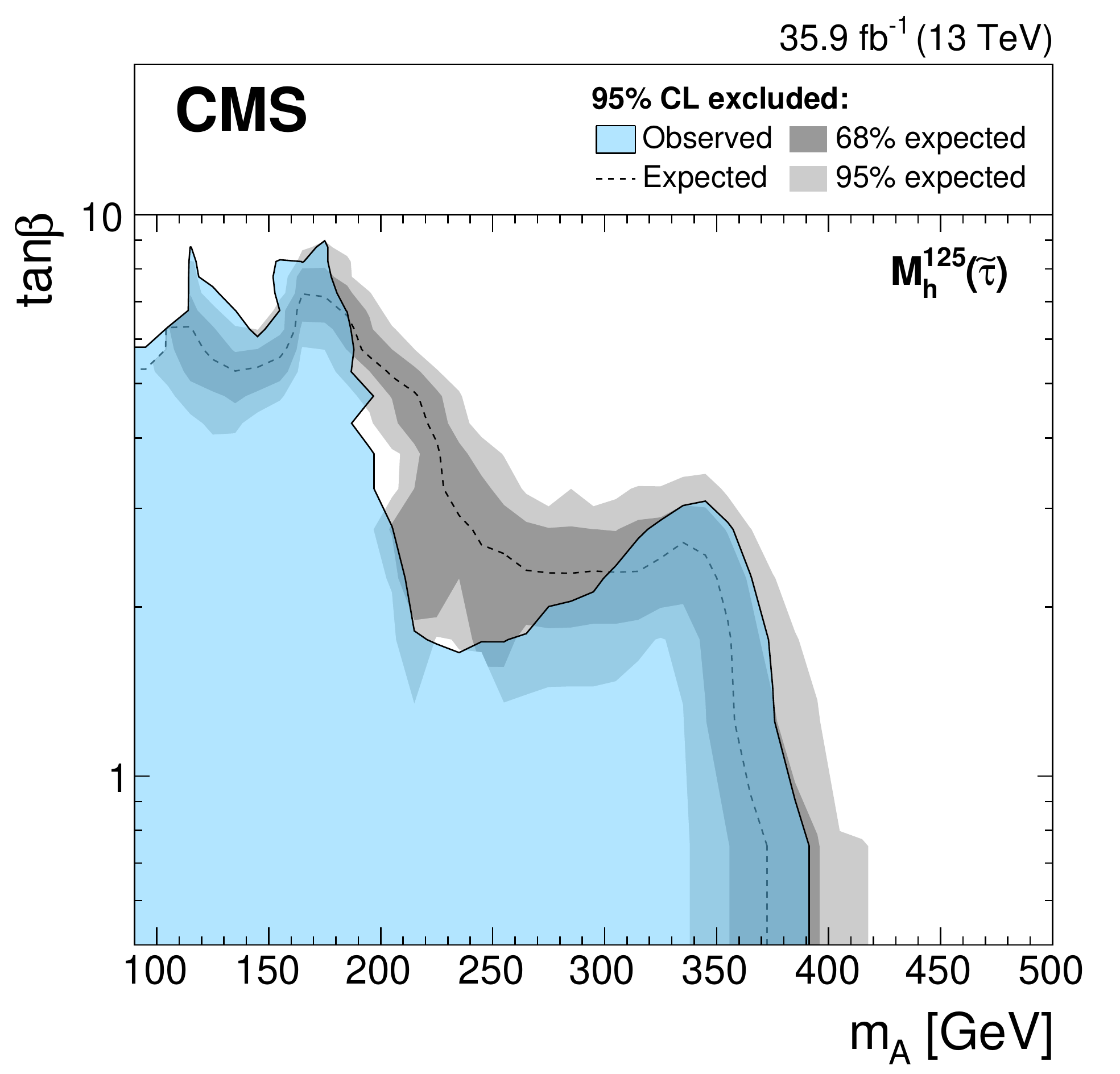} \\
\caption{
Expected and observed 95\% \CL upper limits on \tanbeta as a function of \ma for the \mmodh (upper left), \mmoda (upper right), \mmodc (lower left), and \mmodt (lower right) scenarios.
The expected limit is shown as a dashed black line while the dark
and light gray bands indicate the 68 and 95\% \CL uncertainties, respectively.
The observed exclusion contour is indicated by the blue area.
}
    \label{fig:MSSM2}
\end{figure}

Exclusion limits are also set for neutral heavy Higgs bosons in the context of a Type-\RNum{1} and Type-\RNum{2} 2HDM, with the assumptions that $\mH = \ma = \mHpm$ and $\cosba = 0.1$.
Fig.~\ref{fig:2HDM} shows the expected and observed exclusion limits in the \mH-\tanbeta plane.
The dashed lines mark the expected limits while the dark and bright gray bands indicate the 68 and 95\% \CL uncertainties, respectively.
The observed exclusion contours are indicated by the blue areas.
In both scenarios, the observed exclusion contours reach \mH values of ${\approx}800\GeV$, while the maximum \tanb value excluded is ${\approx}3$.
Fig.~\ref{fig:MSSM} shows the expected and observed exclusion limits for the \mmodp and the hMSSM scenarios.
The maximum \tanb value excluded for both scenarios is ${\approx}9$, while the maximum value of \ma excluded is ${\approx}430\GeV$.
The exclusion of the regions at low values of \ma and \tanbeta complement the exclusion limits set by the MSSM \PH \ra \tautau analyses from ATLAS and CMS using 13\TeV data~\cite{Aaboud:2016cre, Sirunyan:2018zut}, which have reduced sensitivity in these regions.
Fig.~\ref{fig:MSSM2} shows the expected and observed exclusion limits for the \mmodh, \mmoda, \mmodc, and \mmodt scenarios.
Low values of \ma and \tanbeta are also excluded for these scenarios.
The observed exclusion contours reach \ma values of ${\approx}400\GeV$, while the maximum \tanb values excluded are in the range 5--9.
These results further reduce the allowed parameter space for extensions of the SM.

\section{Summary}\label{sec:conclusions}

A search for a heavy Higgs boson decaying to a pair of \W bosons in the mass range from 0.2 to 3.0\TeV has been presented.
The data analysed were collected by the CMS experiment at the LHC in 2016, corresponding to an integrated luminosity of 35.9\fbinv at $\sqrt{s} = 13\TeV$.
The \W boson pair decays are reconstructed in the \FL and \SL final states.
Both gluon fusion and vector boson fusion production of the signal are considered, with a number of hypotheses for their relative contributions investigated.
Interference effects between the signal and background are also taken into account.
Dedicated event categorizations based on both the kinematic properties of associated jets and matrix element techniques are employed to optimize the signal sensitivity.
No evidence for an excess of events with respect to the standard model (SM) predictions is observed.
Combined upper limits at 95\% confidence level on the product of the cross section and branching fraction exclude a heavy Higgs boson with SM-like couplings and decays up to 1870\GeV.
Exclusion limits are also set in the context of a number of two-Higgs-doublet model formulations, further reducing the allowed parameter space for extensions of the SM.

\begin{acknowledgments}\label{sec:acknowledge}
We congratulate our colleagues in the CERN accelerator departments for the excellent performance of the LHC and thank the technical and administrative staffs at CERN and at other CMS institutes for their contributions to the success of the CMS effort. In addition, we gratefully acknowledge the computing centres and personnel of the Worldwide LHC Computing Grid for delivering so effectively the computing infrastructure essential to our analyses. Finally, we acknowledge the enduring support for the construction and operation of the LHC and the CMS detector provided by the following funding agencies: BMBWF and FWF (Austria); FNRS and FWO (Belgium); CNPq, CAPES, FAPERJ, FAPERGS, and FAPESP (Brazil); MES (Bulgaria); CERN; CAS, MoST, and NSFC (China); COLCIENCIAS (Colombia); MSES and CSF (Croatia); RPF (Cyprus); SENESCYT (Ecuador); MoER, ERC IUT, PUT and ERDF (Estonia); Academy of Finland, MEC, and HIP (Finland); CEA and CNRS/IN2P3 (France); BMBF, DFG, and HGF (Germany); GSRT (Greece); NKFIA (Hungary); DAE and DST (India); IPM (Iran); SFI (Ireland); INFN (Italy); MSIP and NRF (Republic of Korea); MES (Latvia); LAS (Lithuania); MOE and UM (Malaysia); BUAP, CINVESTAV, CONACYT, LNS, SEP, and UASLP-FAI (Mexico); MOS (Montenegro); MBIE (New Zealand); PAEC (Pakistan); MSHE and NSC (Poland); FCT (Portugal); JINR (Dubna); MON, RosAtom, RAS, RFBR, and NRC KI (Russia); MESTD (Serbia); SEIDI, CPAN, PCTI, and FEDER (Spain); MOSTR (Sri Lanka); Swiss Funding Agencies (Switzerland); MST (Taipei); ThEPCenter, IPST, STAR, and NSTDA (Thailand); TUBITAK and TAEK (Turkey); NASU (Ukraine); STFC (United Kingdom); DOE and NSF (USA).

\hyphenation{Rachada-pisek} Individuals have received support from the Marie-Curie programme and the European Research Council and Horizon 2020 Grant, contract Nos.\ 675440, 752730, and 765710 (European Union); the Leventis Foundation; the A.P.\ Sloan Foundation; the Alexander von Humboldt Foundation; the Belgian Federal Science Policy Office; the Fonds pour la Formation \`a la Recherche dans l'Industrie et dans l'Agriculture (FRIA-Belgium); the Agentschap voor Innovatie door Wetenschap en Technologie (IWT-Belgium); the F.R.S.-FNRS and FWO (Belgium) under the ``Excellence of Science -- EOS" -- be.h project n.\ 30820817; the Beijing Municipal Science \& Technology Commission, No. Z181100004218003; the Ministry of Education, Youth and Sports (MEYS) of the Czech Republic; the Deutsche Forschungsgemeinschaft (DFG) under Germany’s Excellence Strategy -- EXC 2121 ``Quantum Universe" -- 390833306; the Lend\"ulet (``Momentum") Programme and the J\'anos Bolyai Research Scholarship of the Hungarian Academy of Sciences, the New National Excellence Program \'UNKP, the NKFIA research grants 123842, 123959, 124845, 124850, 125105, 128713, 128786, and 129058 (Hungary); the Council of Science and Industrial Research, India; the HOMING PLUS programme of the Foundation for Polish Science, cofinanced from European Union, Regional Development Fund, the Mobility Plus programme of the Ministry of Science and Higher Education, the National Science Center (Poland), contracts Harmonia 2014/14/M/ST2/00428, Opus 2014/13/B/ST2/02543, 2014/15/B/ST2/03998, and 2015/19/B/ST2/02861, Sonata-bis 2012/07/E/ST2/01406; the National Priorities Research Program by Qatar National Research Fund; the Ministry of Science and Education, grant no. 3.2989.2017 (Russia); the Programa Estatal de Fomento de la Investigaci{\'o}n Cient{\'i}fica y T{\'e}cnica de Excelencia Mar\'{\i}a de Maeztu, grant MDM-2015-0509 and the Programa Severo Ochoa del Principado de Asturias; the Thalis and Aristeia programmes cofinanced by EU-ESF and the Greek NSRF; the Rachadapisek Sompot Fund for Postdoctoral Fellowship, Chulalongkorn University and the Chulalongkorn Academic into Its 2nd Century Project Advancement Project (Thailand); the Nvidia Corporation; the Welch Foundation, contract C-1845; and the Weston Havens Foundation (USA).
\end{acknowledgments}

\bibliography{auto_generated}

\cleardoublepage \appendix\section{The CMS Collaboration \label{app:collab}}\begin{sloppypar}\hyphenpenalty=5000\widowpenalty=500\clubpenalty=5000\input{HIG-17-033-authorlist.tex}\end{sloppypar}
\end{document}

%% file: HIG-17-033-authorlist.tex
\vskip\cmsinstskip
\textbf{Yerevan Physics Institute, Yerevan, Armenia}\\*[0pt]
A.M.~Sirunyan$^{\textrm{\dag}}$, A.~Tumasyan
\vskip\cmsinstskip
\textbf{Institut f\"{u}r Hochenergiephysik, Wien, Austria}\\*[0pt]
W.~Adam, F.~Ambrogi, T.~Bergauer, J.~Brandstetter, M.~Dragicevic, J.~Er\"{o}, A.~Escalante~Del~Valle, M.~Flechl, R.~Fr\"{u}hwirth\cmsAuthorMark{1}, M.~Jeitler\cmsAuthorMark{1}, N.~Krammer, I.~Kr\"{a}tschmer, D.~Liko, T.~Madlener, I.~Mikulec, N.~Rad, J.~Schieck\cmsAuthorMark{1}, R.~Sch\"{o}fbeck, M.~Spanring, D.~Spitzbart, W.~Waltenberger, C.-E.~Wulz\cmsAuthorMark{1}, M.~Zarucki
% ...
\vskip\cmsinstskip
\dag: Deceased\\
1:  Also at Vienna University of Technology, Vienna, Austria\\
% ...
